\documentclass[aps,floats,twocolumn,superscriptaddress]{revtex4}
\usepackage[utf8]{inputenc}  
\usepackage[T1]{fontenc} 
\usepackage[squaren,Gray]{SIunits}
\usepackage{graphicx}
\usepackage{subfigure}
\DeclareGraphicsExtensions{.eps,.ps,.pdf}
\usepackage{amsmath,amssymb,bbold,bm}
\usepackage[pdfusetitle,bookmarksnumbered,plainpages=false]{hyperref}
\usepackage{xcolor}
\usepackage{comment}
\usepackage[english]{babel}

\begin{document}


\newcommand{\etal}{\emph{et al.}}
\newcommand{\LCO}{$\rm La_2CuO_4$}
\newcommand{\CaNCOCl}{$\rm Ca_{2-x}Na_xCuO_2Cl_2$}
\newcommand{\Biddzu}{$\rm Bi_2Sr_{2-x}La_xCuO_6$}
\newcommand{\Biddud}{$\rm Bi_2Sr_2CaCu_2O_{8+\delta}$}
\newcommand{\PCCO}{$\rm Pr_{2-x}Ce_xCuO_{4-\delta}$}
\newcommand{\LSCOud}{La$_{1.855}$Sr$_{0.145}$CuO$_4$ }
\newcommand{\LSCOov}{La$_{1.83}$Sr$_{0.17}$CuO$_4$ }
\newcommand{\LSCO}{La$_{2-x}$Sr$_x$CuO$_4$}
\newcommand{\LSCOten}{La$_{1.90}$Sr$_{0.10}$CuO$_4$}
\newcommand{\LSCOtwelve}{La$_{1.88}$Sr$_{0.12}$CuO$_4$}
\newcommand{\LSCOfourteen}{La$_{1.86}$Sr$_{0.14}$CuO$_4$}
\newcommand{\LSCOsuper}{La$_{2-x}$Sr$_x$CuO$_{4 + y}$}
\newcommand{\NDLSCO}{La$_{2-x-y}$Nd$_y$Sr$_x$CuO$_4$}
\newcommand{\NDLSCOtwelve}{La$_{1.48}$Nd$_{0.4}$Sr$_{0.12}$CuO$_4$}
\newcommand{\NDLSCOft}{La$_{1.45}$Nd$_{0.4}$Sr$_{0.15}$CuO$_4$}
\newcommand{\LBCO}{La$_{2-x}$Ba$_x$CuO$_4$}
\newcommand{\HgCO}{$\rm HgBa_2Ca_2Cu_3O_{8+x}$}
\newcommand{\Hgudzu}{$\rm HgBa_2CuO_{4+\delta}$}
\newcommand{\Tlddzu}{$\rm Tl_2Ba_2CuO_{6+\delta}$}
\newcommand{\YBCO}{YBa$_2$CuO$_{7+\delta}$}
\newcommand{\Tc}{$T_{\rm c}$}
\newcommand {\CuOd}{$\rm CuO_{\rm 2}$}
\newcommand{\Trho}{$T_{\rho}$}
\newcommand{\Tst}{$T^*$}
\newcommand{\Tz}{$T_0$}
\newcommand{\Tzr}{$T_0^\rho$}
\newcommand{\Tmax}{$T_{\rm max}$}
\newcommand{\rh}{$R_{\rm H}$}
\newcommand{\Toxy}{$T_{\rm oxygene}$}
\newcommand{\Tnu}{$T_\nu$}
\newcommand{\Tmr}{$T_{\rm MR}$}
\newcommand{\rhoch}{$\rho_{\rm chaine}$}
\newcommand{\Yquatr}{$\rm YBa_2Cu_4O_8$}
\newcommand{\Ytroi}{$\rm YBa_2Cu_3O_y$}
\newcommand{\Ysz}{$\rm YBa_2Cu_3O_{6}$}
\newcommand{\Ytc}{$\rm YBa_2Cu_3O_{6.35}$}
\newcommand{\Yqc}{$\rm YBa_2Cu_3O_{6.45}$}
\newcommand{\Yqn}{$\rm YBa_2Cu_3O_{6.49}$}
\newcommand{\Ycu}{$\rm YBa_2Cu_3O_{6.51}$}
\newcommand{\Ycq}{$\rm YBa_2Cu_3O_{6.54}$}
\newcommand{\Yss}{$\rm YBa_2Cu_3O_{6.67}$}
\newcommand{\Ysq}{$\rm YBa_2Cu_3O_{6.64}$}
\newcommand{\Ysc}{$\rm YBa_2Cu_3O_{6.75}$}
\newcommand{\Yhz}{$\rm YBa_2Cu_3O_{6.80}$}
\newcommand{\Yhs}{$\rm YBa_2Cu_3O_{6.86}$}
\newcommand{\Ynd}{$\rm YBa_2Cu_3O_{6.92}$}
\newcommand{\Yns}{$\rm YBa_2Cu_3O_{6.97}$}
\newcommand{\Ynnh}{$\rm YBa_2Cu_3O_{6.998}$}
\newcommand{\Ync}{$\rm YBa_2Cu_3O_{6.95}$}
\newcommand{\Ys}{$\rm YBa_2Cu_3O_{7}$}
\newcommand{\CaY}{$\rm Ca_xY_{1-x}Ba_2Cu_3O_{6.99}$}
\newcommand{\NbSet}{$\rm NbSe_3$}
\newcommand{\NbSed}{$\rm NbSe_2$}
\newcommand{\SHTC}{supraconductivit� � haute $T_{\rm c}$}
\newcommand{\Hirr}{$H_{\rm irr}$}
\newcommand{\Hc}{$H_{\rm c2}$}
\newcommand{\Hco}{$H_{\rm CO}$}
\newcommand{\Tco}{$T_{\rm CO}$}
\newcommand{\scap}{\!\cdot\!}
\newcommand{\intd}[1]{{\int\!d#1\: }}
\newcommand{\inttau}[1]{{\int_0^\beta\!d\tauno_{#1}\: }}
\newcommand{\dpst}{\displaystyle}
\newcommand{\scst}{\scriptstyle}
\newcommand{\scscst}{\scriptscriptstyle}
\newcommand{\Bv}{$\mathbf{B}$}

\newcommand{\beg}{\begin{equation}}
\newcommand{\eeq}{\end{equation}}
\newcommand{\beqa}{\begin{eqnarray}}
\newcommand{\eeqa}{\end{eqnarray}}

\title{High magnetic field ultrasound study of spin freezing in La$_{1.88}$Sr$_{0.12}$CuO$_4$}

\author{M. Frachet}
\thanks{mehdi.frachet@kit.edu. Present adress : IQMT, Karlsruhe Institute für Technologie, 76021 Karlsruhe, Germany}
\affiliation{Univ. Grenoble Alpes, INSA Toulouse, Univ. Toulouse Paul Sabatier, EMFL, CNRS, LNCMI, 38000, Grenoble, France}
\author{S. Benhabib}
\affiliation{Univ. Grenoble Alpes, INSA Toulouse, Univ. Toulouse Paul Sabatier, EMFL, CNRS, LNCMI, 38000, Grenoble, France}
\author{I. Vinograd}
\affiliation{Univ. Grenoble Alpes, INSA Toulouse, Univ. Toulouse Paul Sabatier, EMFL, CNRS, LNCMI, 38000, Grenoble, France}
\author{S.-F. Wu}
\affiliation{Univ. Grenoble Alpes, INSA Toulouse, Univ. Toulouse Paul Sabatier, EMFL, CNRS, LNCMI, 38000, Grenoble, France}
\author{B. Vignolle}
\affiliation{Institut de Chimie de la Matière Condensée, Bordeaux, France}
\author{H. Mayaffre}
\affiliation{Univ. Grenoble Alpes, INSA Toulouse, Univ. Toulouse Paul Sabatier, EMFL, CNRS, LNCMI, 38000, Grenoble, France}
\author{S. Krämer}
\affiliation{Univ. Grenoble Alpes, INSA Toulouse, Univ. Toulouse Paul Sabatier, EMFL, CNRS, LNCMI, 38000, Grenoble, France}
\author{T. Kurosawa}
\affiliation{Department of Physics, Hokkaido University, Sapporo 060-0810, Japan}
\author{N. Momono}
\affiliation{Muroran Institute of Technology, Muroran 050-8585, Japan}
\author{M. Oda}
\affiliation{Department of Physics, Hokkaido University, Sapporo 060-0810, Japan}
\author{J. Chang}
\affiliation{Department of Physics, University of Zurich, CH-8057 Zurich, Switzerland}
\author{C. Proust}
\affiliation{Univ. Grenoble Alpes, INSA Toulouse, Univ. Toulouse Paul Sabatier, EMFL, CNRS, LNCMI, 38000, Grenoble, France}
\author{M.-H. Julien}
\affiliation{Univ. Grenoble Alpes, INSA Toulouse, Univ. Toulouse Paul Sabatier, EMFL, CNRS, LNCMI, 38000, Grenoble, France}
\author{D. LeBoeuf} 
\thanks{david.leboeuf@lncmi.cnrs.fr}
\affiliation{Univ. Grenoble Alpes, INSA Toulouse, Univ. Toulouse Paul Sabatier, EMFL, CNRS, LNCMI, 38000, Grenoble, France}

\begin{abstract}

High-$T_{\rm{c}}$ cuprate superconductors host spin, charge and lattice instabilities. In particular, in the antiferromagnetic glass phase, over a large doping range, lanthanum based cuprates display a glass-like spin freezing with antiferromagnetic correlations. Previously, sound velocity anomalies in La$_{2-x}$Sr$_{x}$CuO$_4$ (LSCO) for hole doping $p\geq 0.145$ were reported and interpreted as arising from a coupling of the lattice to the magnetic glass [Frachet, Vinograd \emph{et al.}, Nat. Phys. \textbf{16}, 1064-1068 (2020)]. Here we report both sound velocity and attenuation in LSCO $p=0.12$, \textit{i.e.} at a doping level for which the spin freezing temperature is the highest. Using high magnetic fields and comparing with nuclear magnetic resonance (NMR) measurements, we confirm that the anomalies in the low temperature ultrasound properties of LSCO are produced by a coupling between the lattice and the spin glass. Moreover, we show that both sound velocity and attenuation can be simultaneously accounted for by a simple phenomenological model originally developed for canonical spin glasses. Our results point towards a strong competition between superconductivity and spin freezing, tuned by the magnetic field. A comparison of different acoustic modes suggests that the slow spin fluctuations have a nematic character.

\end{abstract}

\date{\today}
\maketitle

\section{Introduction}
The coupling of electronic instabilities to the crystal lattice plays a significant role in shaping the phase diagram of some high-$T_{\rm c}$ cuprate superconductors. The case of La-based cuprates is emblematic. Upon cooling, La$_{2-x}$Ba$_{x}$CuO$_4$ (LBCO) and rare-earth doped (Nd, Eu)$_{y}$-La$_{2-x-y}$Sr$_{x}$CuO$_4$ ((Nd,Eu)-LSCO) evolve from a high-$T$ tetragonal (HTT) to a mid-$T$ orthorhombic (OMT) and finally to a low-$T$ tetragonal (LTT) crystal structure. The LTT order pins stripe order, a combination of mutually commensurate spin and charge modulations, initially found in Nd-LSCO \citep{Tranquada_Nature_1995}. Within this context sound velocity and attenuation are particularly relevant quantities. Ultrasound measurements directly probe the lattice properties and they are sensitive to any strain dependent instability.


Among the La-based cuprate family La$_{2-x}$Sr$_{x}$CuO$_4$ (LSCO) appears peculiar. First, the OMT-LTT structural phase transition does not occur, although LTT-like distortions exist locally \citep{Bozin_PRB_99, Bianconi_PRL_1996, Saini_PRB_1997}. 
Moreover, scattering evidence for charge ordering inside the pseudogap phase has remained elusive until recently \citep{Thampy_PRB_14, Croft_PRB_14,Christensen_14,Wen_Arxiv_CDW_LSCO}. 
In La$_{1.88}$Sr$_{0.12}$CuO$_4$ quasi-static charge modulation appears below $T_{\rm{CDW}}=70 \pm 15$ K with a maximal in-plane correlation length $\xi_{\parallel}(T_{\rm{c}}) \simeq 30~\angstrom$, a value practically one order of magnitude smaller than in LBCO at the same doping. 

In the same compound incommensurate antiferromagnetic (AFM) correlations are also found at low field for $0.02 \leq p \lesssim 0.135$ \citep{MHJ_review_LSCO}. The temperature at which these correlations appear static depends upon the probe frequency \citep{MHJ_review_LSCO, Romer_PRB_13}, revealing the glassy nature of the magnetic state. However, as in other La-based compounds close to $p \approx 0.12$, one observes that the incommensurabilities of charge and spin density waves (respectively CDW and SDW) follow $2\delta_{\rm{spin}}=\delta_{\rm{charge}}$, a relation reminiscent of charge-spin stripe ordering \citep{Croft_PRB_14}. 

Close to the hole doping level $p \approx 0.12$~elastic anomalies have been reported in both sound velocity and attenuation. Specifically, in single crystal studies and near the superconducting $T_{\rm c}$, a broad sound velocity minimum has been observed in different acoustic modes \citep{Suzuki_PRB_98, Nohara_PRL_93}. In a similar range of temperature, an attenuation maximum of longitudinal waves has been found in polycrystals \citep{Qu_APL_2006, Qu_MSE_2006}. Different interpretations have been proposed to explain this peculiar behaviour \citep{Nohara_PRL_93,Qu_APL_2006,Hanaguri_PhysicaB_94, Sakita_PhysicaB_96, Suzuki_PRB_98, Prieur_04}. Recently, using NMR and sound velocity measurements in high magnetic field in LSCO for $p \geq 0.145$, we have shown that the anomalous sound velocity appears to be caused by a coupling of the AFM glass to the lattice \citep{Frachet_19}.

In this study, we strengthen this interpretation with high magnetic field measurements of sound velocity and attenuation in LSCO $p=0.12$. Comparing ultrasound attenuation with NMR measurements on crystals from the same batch, we reinforce the link between the slowing down of magnetic fluctuations and the ultrasound anomalies observed in the $(c_{11}-c_{12})/2$ and $c_{11}$ elastic constants. Moreover, we show that the ultrasound properties of the $(c_{11}-c_{12})/2$ mode can be semi-quantitatively reproduced by a phenomenological dynamical susceptibility model initially developed for canonical spin glasses. Finally, by comparing different acoustic modes, we find that the spin freezing produces an enhanced susceptibility in the $B_{\rm{1g}}$ channel, which is associated with nematicity in cuprates.

This paper is organized as follows. In section II, we describe the sample studied and the experimental technique. Then, in section III we report the experimental sound velocity and attenuation measurements. We present a phenomenological model of ultrasound in spin glasses and use it to analyze the ultrasound data in section IV. Then, in section V, we discuss the magnetic field effect on the ultrasound properties, the differences between the acoustic modes studied and the symmetry of the AFM fluctuations inferred from our measurements. We summarize our conclusions in section VI.

\section{Methods}
A high quality LSCO single crystal was grown by the traveling solvent floating zone method. From this crystal, three samples were cut along different crystallographic directions in order to probe different elastic constants. Typical samples dimensions are $2 \times 2 \times 2$ mm$^3$. The hole doping $p = 0.122 \pm 0.002$ has been determined by measuring $T_{\rm{st}} = 252 ~\rm{K}$, the temperature of the HTT-OMT structural phase transition by sound velocity, as described in Ref. \citep{Frachet_19}. The different samples share a similar $T_{\rm{st}}$ and thus a similar doping. The superconducting transition temperature $T_{\rm{c}} = 29 \pm 3$ K has been determined by sound velocity, in-plane resistivity, and magnetic susceptibility measurements. 

A standard pulse-echo technique with phase comparison was used to measure variations of sound velocity, $\Delta v/v$, and sound attenuation, $\Delta \alpha$ \citep{Luthi_book}. Ultrasound was generated and detected using commercial LiNbO$_3$ transducers glued onto parallel, clean and polished surfaces of the samples. The excitation frequency, $\omega$, ranged from 50 to 300 MHz. For high symmetry propagation direction, the sound velocity variation of a given acoustic mode can be converted to the associated elastic constant change using $\Delta c_{\rm{ii}}/c_{\rm{ii}} = 2 \Delta v/v$. 

Zero-field and static-field experiments were performed at the LNCMI Grenoble using 20 T superconducting and 28 T resistive magnets. Field cooled conditions were used. Pulsed-fields experiments up to 60 T were carried out at the LNCMI Toulouse. In all cases, the field was applied along the crystallographic $c$-axis.

\section{Results}

\subsection{Sound velocity in zero magnetic field}

\begin{figure}
\includegraphics[width=8.6cm]{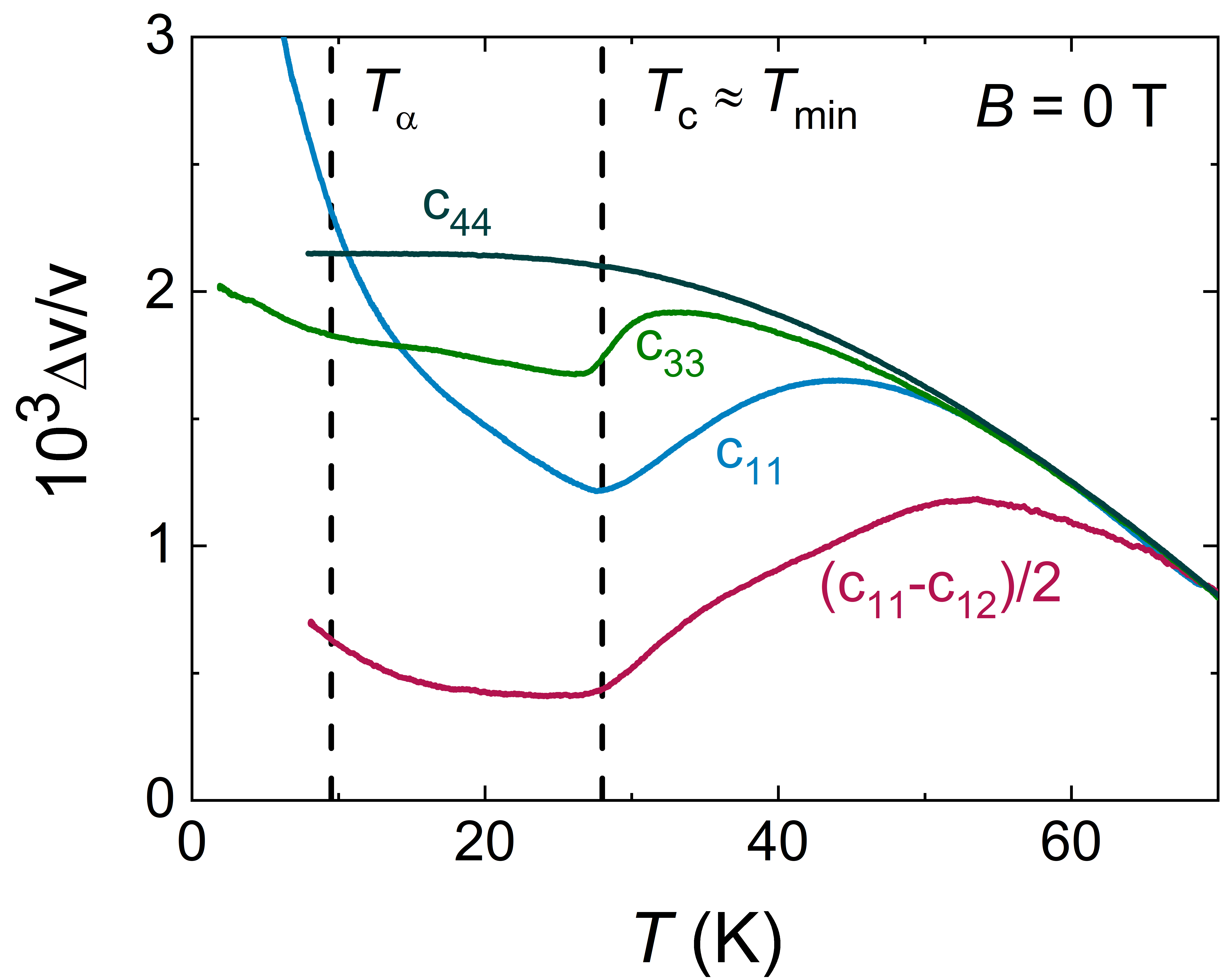}
\caption{Sound velocity variation, $\Delta v/v$, as a function of temperature for different acoustic modes as indicated. No magnetic field is applied, the curves are arbitrarily superimposed at \textit{T}=70 K. \textit{T}$_{\rm{min}}$ refers to the minimum in the $c_{11}$ elastic constant that coincides with the superconducting $T_{\rm{c}}$ in zero magnetic field. $T_{\rm{\alpha}}~$ indicates the spin freezing temperature at the $\mu$eV ultrasound energy scale, defined by an attenuation peak (see Fig. \ref{fig2}).}
\label{fig1}
\end{figure}

\begin{figure*}
\centering
\includegraphics[width=17.2cm]{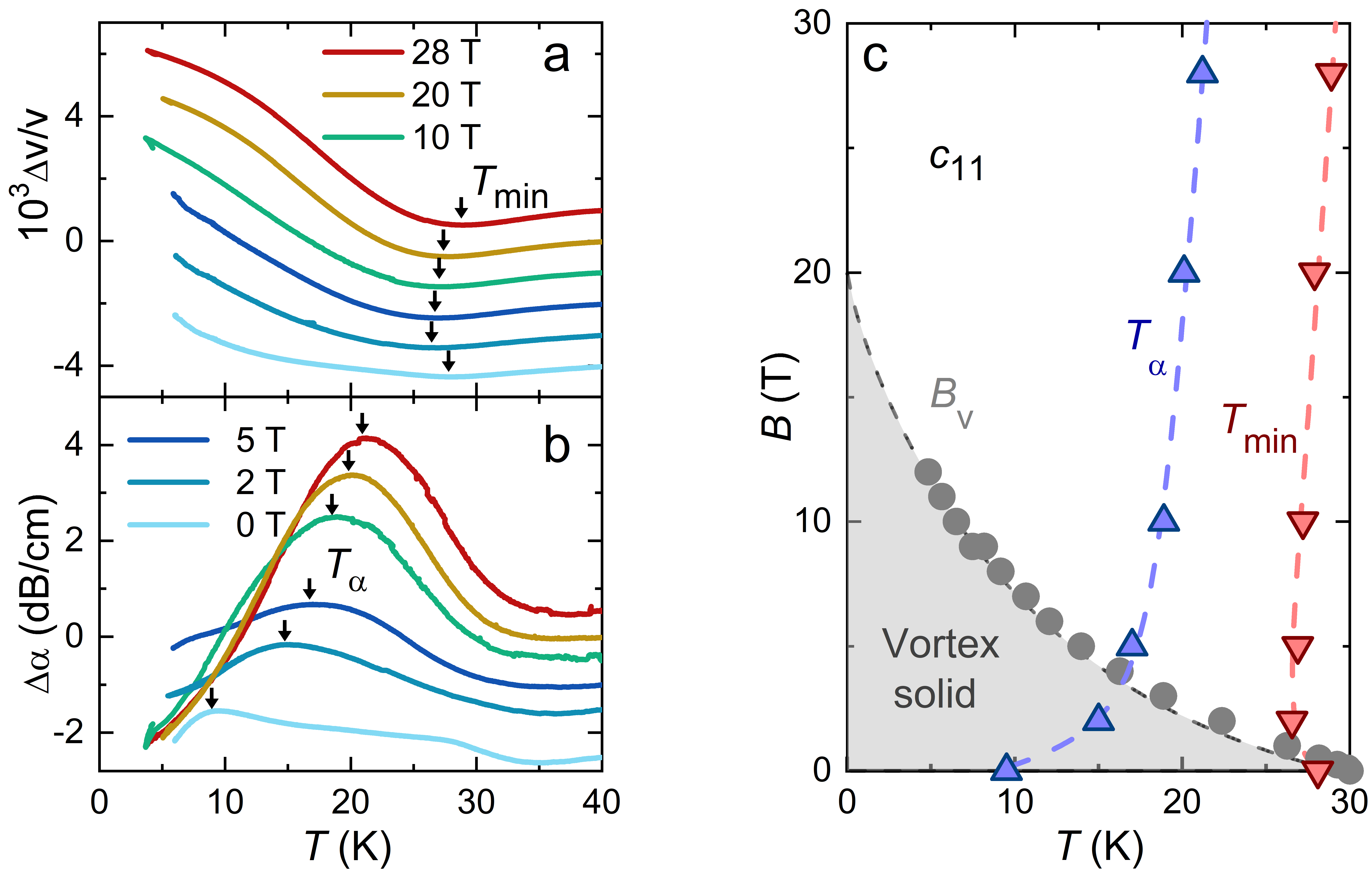}
\caption{Static field ultrasound measurements in the $c_{11}$ acoustic mode up to 28 T. (a) $\Delta v/v$ and (b) $\Delta \alpha$ as a function of temperature for different magnetic fields. The curves are shifted for clarity. The arrows denote $T_{\rm{min}}$ and $T_{\rm{\alpha}}$ as indicated. (c) $B - T$ phase diagram: gray circles denote the vortex melting transition field $B_{\rm v}$ inferred from in-plane resistivity measurements, $T_{\rm{\alpha}}$ (upward-pointing blue triangles) and $T_{\rm{min}}$ (downward-pointing red triangles) refer to the maximum in attenuation and minimum in velocity respectively. Error bars are smaller than the symbol size. Dashed lines are guides to the eye.}
\label{fig2}
\end{figure*}

\begin{table}
\begin{tabular}{c || c | c | c | c}
\textbf{$c_{ij}$} & \textbf{$\vec{k}$} & \textbf{$\vec{u}$} &  $\epsilon_{ij}$ & symmetry\\
  \hline
  \hline
$c_{11}$ &     [100]   & [100] & $\epsilon_{xx}$ & $A_{1g} + B_{1g}$\\
  \hline
$c_{33}$ & [001] & [001] & $\epsilon_{zz}$ & $A_{1g}$\\
  \hline
$c_{44}$ & [100] & [001] & $\epsilon_{yz}$, $\epsilon_{zx}$ & $E_g$\\  
  \hline
$(c_{11}-c_{12})/2 $ & [110] & [1$\overline{1}$0] & $\epsilon_{xx} - \epsilon_{yy}$ & $B_{1g}$\\ 
  
\end{tabular}
\caption{Properties of the different elastic constants measured: direction of propagation ($\vec{k}$) and polarisation ($\vec{u}$) of the acoustic wave, strain ($\epsilon_{ij}$) and associated symmetry. The indices of elastic constants are expressed in the Voigt notation. Crystallographic directions are those of the HTT phase ($D_{4h}$ point group).}
\label{table_elastic_constant}
\end{table}

We begin with a zero magnetic field study of different elastic constants in LSCO $p=0.12$ as shown in Fig. \ref{fig1}. The description of the different modes studied is reported in table \ref{table_elastic_constant}. The $c_{44}$ acoustic mode follows the classical variation expected in solids: upon cooling the sound velocity increases continuously and eventually saturates at low temperature \citep{Varshni_PRB_70}. 
This behaviour contrasts with the $c_{33}$ elastic constant which shows a downward jump at $T_{\rm{c}}$. This mean-field anomaly at the superconducting transition is expected for a longitudinal mode and is related to the specific heat jump through the Ehrenfest relationship:
\begin{equation}
\Delta c_{ii} (T_{\rm{c}})= - \frac{\Delta C_p(T_{\rm{c}})}{V_{\rm{mol}} T_{\rm{c}} } \left(\frac{d T_{\rm{c}}}{d \epsilon_i} \right)^2
\label{ehrenfest_eq}
\end{equation} with $\Delta C_p(T_{\rm{c}})$ the specific heat jump at $T_{\rm{c}}$ and $V_{\rm{mol}}$ the molar volume. The amplitude of the anomaly, $\Delta v/v (T_{\rm{c}}) \simeq  0.2 \times 10^{-3}$, is consistent with literature values on samples with similar doping levels \citep{Nohara_PRB_95, Gugenberger_PRB_94}.

For $T\geq T_{\rm{c}}$, the temperature dependence of $c_{11}$ and $(c_{11}-c_{12})/2$ elastic constants is anomalous. In both of these modes, the normal state sound velocity decreases upon cooling, until the temperature hits $T_{\rm c}$ where it shows an upturn. Consequently, the sound velocity in these modes has a minimum at $T_{\rm{min}} \simeq T_{\rm{c}}$. Fig. \ref{fig1} shows that the anomalous lattice softening appears only in acoustic modes having a $B_{\rm 1g}$ strain  component, namely $c_{11}$ and $(c_{11}-c_{12})/2$. Note that so far we have not been able to measure the $B_{\rm 2g}$ mode ($c_{66}$) for $T < T_{\rm{st}}$.

Finally, for $T \lesssim 15 ~$K or so, a rapid stiffening is observed in $c_{11}$ upon cooling. Indeed, the sound velocity in the $T=0$ limit exceeds largely what would be expected from an extrapolation of the high temperature bare elastic constant (\textit{e.g.} following the $c_{44}$ elastic constant). A similar upturn is found in $c_{33}$ and $(c_{11}-c_{12})/2$ upon cooling for $T\lesssim 15~$K, although much weaker than in $c_{11}$.


\subsection{Sound velocity and attenuation in applied magnetic field}

\begin{figure}
\centering
\includegraphics[width=8.6cm]{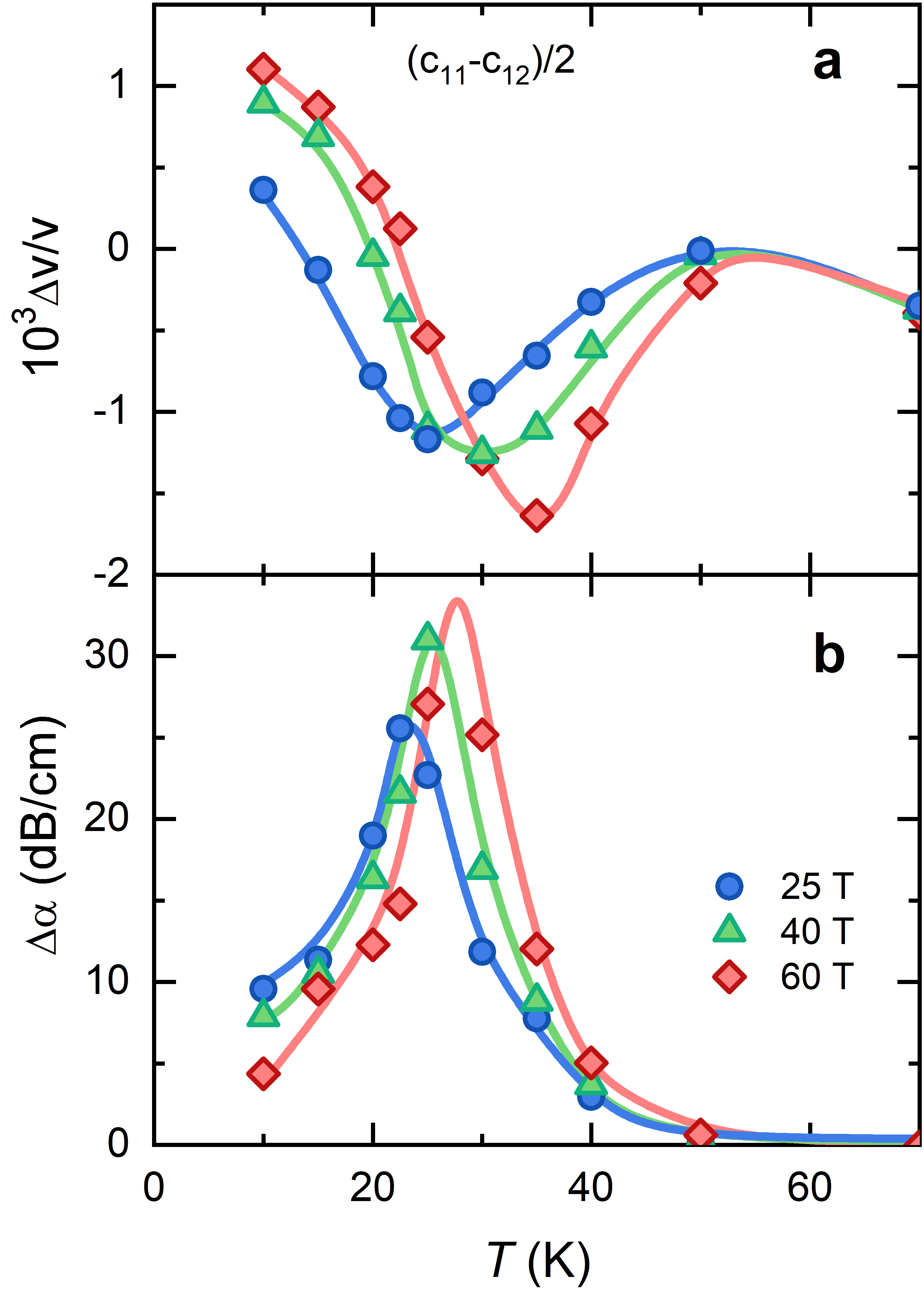}
\caption{Temperature dependence of the $(c_{11}-c_{12})/2$ acoustic mode in pulsed-fields up to 60 T. (a) $\Delta v/v$ and (b) $\Delta \alpha$ both extracted from fixed magnetic field cuts of the pulsed-fields isotherms to which we add the zero field curve. All lines are guides to the eye.}
\label{fig3}
\end{figure}

In Fig. \ref{fig2} and Fig. \ref{fig3} we investigate how the anomalous sound velocity, and the corresponding sound attenuation, evolve as a function of temperature at different magnetic fields, in the $c_{11}$ and  $(c_{11}-c_{12})/2$ modes respectively. In both these modes no signature of the vortex lattice is observed, as discussed in   Appendix \ref{appendix_vortex}.

The anomalous features of the zero field sound velocity in the $c_{11}$ and  $(c_{11}-c_{12})/2$ acoustic modes are enhanced by a magnetic field: both the amplitudes of the lattice softening (for $T \geq T_{\rm{min}}$) and stiffening ($T \leq T_{\rm{min}}$) increase with increasing magnetic field. 
For both acoustic modes an attenuation peak is found at $T_{\rm{\alpha}} \leq T_{\rm{min}}$. The amplitude of this attenuation peak and $T_{\rm{\alpha}}$ increase monotonically with increasing field. 

The magnetic field dependencies of $T_{\rm{\alpha}}$ and $T_{\rm{min}}$  from $c_{11}$ measurements are shown in the phase diagram of Fig. \ref{fig2}(c). Within error bars, $(c_{11}-c_{12})/2$ and $c_{11}$ show at a given magnetic field similar $T_{\rm \alpha}$ and $T_{\rm min}$. In contrast with $T_{\alpha}$, $T_{\rm{min}}$ has a non-monotonic field dependence: it decreases for $0\leq B\leq 2$ T and increases for higher fields. The initial decrease is caused by the lattice coupling to the superconducting order parameter as further detailed in Appendix \ref{appendix_superconductivity}. However, for $B\geq 5$~T or so, the two temperature scales have similar field dependence, indicating that they are coupled and caused by the same phenomenon.

\subsection{Comparison with NMR $1/T_1$}
\begin{figure}
\centering
\includegraphics[width=8.6cm]{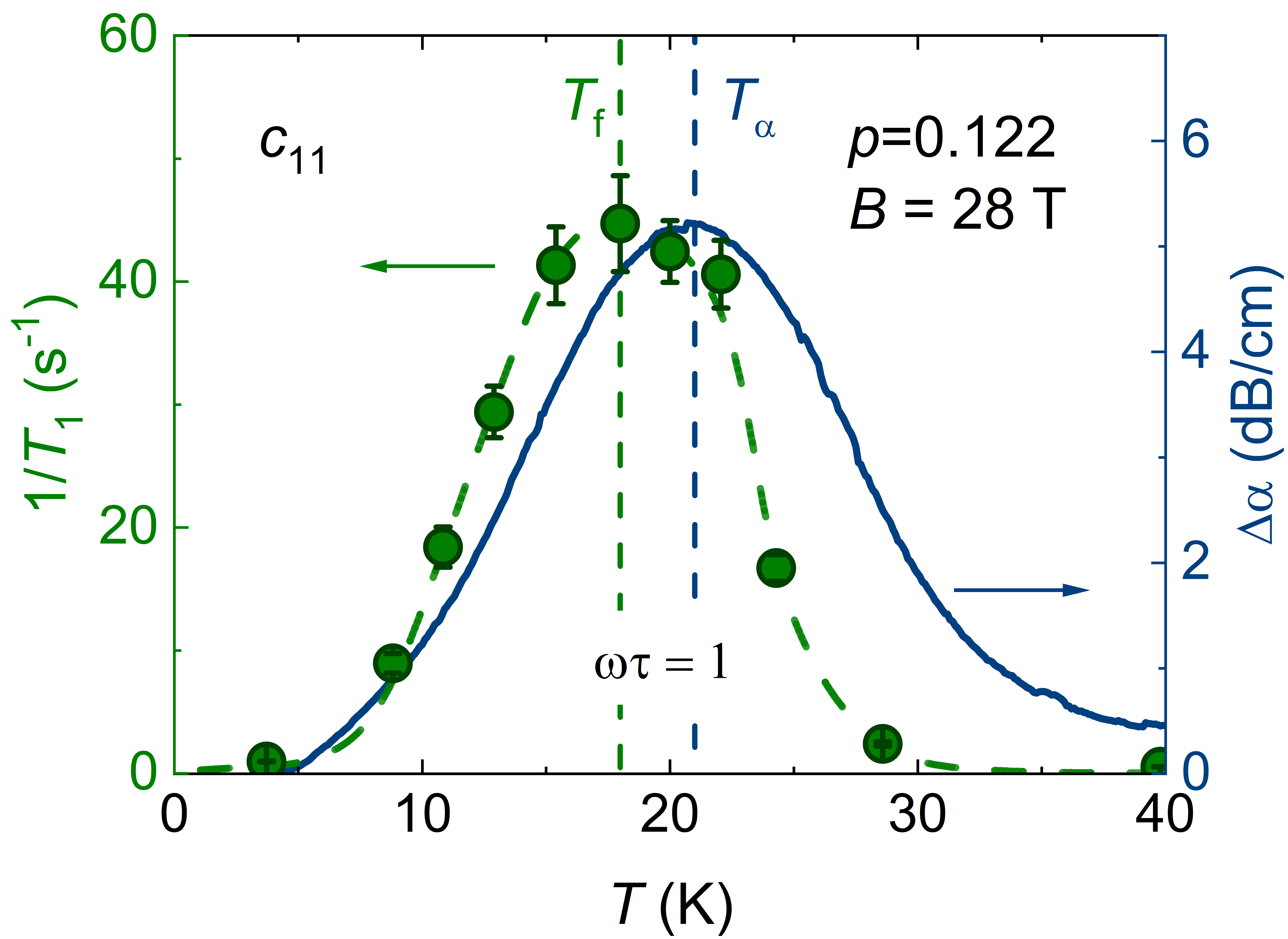}
\caption{Comparison between the ultrasound attenuation $\Delta \alpha$ (blue line, right scale) and the $^{139}$La NMR spin-lattice relaxation rate $1/T_1$ (green circle, left scale), at $B = 28$ T. NMR and ultrasound samples are from the same batch. Both physical quantities show a peak when the excitation frequency becomes equal to the frequency of the spin fluctuations, $i.e.$ $\omega\tau=1$. A phenomenological linear background has been substracted from the experimental $\Delta \alpha$ for clarity. The dashed green line is a guide to the eye.}
\label{fig4}
\end{figure}

In Fig. \ref{fig4} we compare the ultrasound attenuation, $\Delta \alpha$, with the $^{139}$La NMR spin-lattice relaxation rate, $1/T_1$, both measured in LSCO $p=0.12$ samples from the same batch and in a magnetic field $B=28~$T. The comparison is striking, both quantities display remarkably similar temperature dependencies and show a maximum at comparable temperatures.

The peak in $1/T_1$ is a classical signature of spin freezing in superconducting LSCO \citep{Mitrovic_PRB_08, Hunt_PRB_01, Simovic_PRB_03,Curro_PRL_00}. This peak is understood within the so-called Bloembergen-Purcell-Pound (BPP) model \citep{BPP_PR_1948, Mitrovic_PRB_08, Curro_PRL_00} originating from a diverging correlation time $\tau(T)$. Upon cooling, spin fluctuations are gradually slowing down. At the temperature where the condition $\omega_{\rm NMR} \tau=1$ is fulfilled ($\omega_{\rm NMR}$ being the NMR frequency), $1/T_1$ is maximum. This temperature defines the freezing temperature $T_{\rm f}$ at the NMR time-scale.

Correspondingly, Fig. \ref{fig4} reveals that the ultrasound attenuation $\Delta \alpha$ is governed by a similar correlation time. At the temperature where the condition $\omega_{\rm US} \tau=1$ is met, with $\omega_{\rm US}$ the ultrasound frequency, a peak in the ultrasound attenuation is observed. The good agreement between $T_{\rm f}$ and $T_{\rm \alpha}$ is provided by the fact that $\omega_{\rm NMR} \approx \omega_{\rm US} \approx 10^8$ Hz. Finally, note that the small difference observed between $\Delta \alpha$ and $1/T_1$ could arise from a small variation in doping level between the two samples, but also from a disparity in the way these probes couple to the magnetic moments. This is discussed in the next section.

\section{Modelling}

\begin{figure*}
\centering
\includegraphics[width=17.2cm]{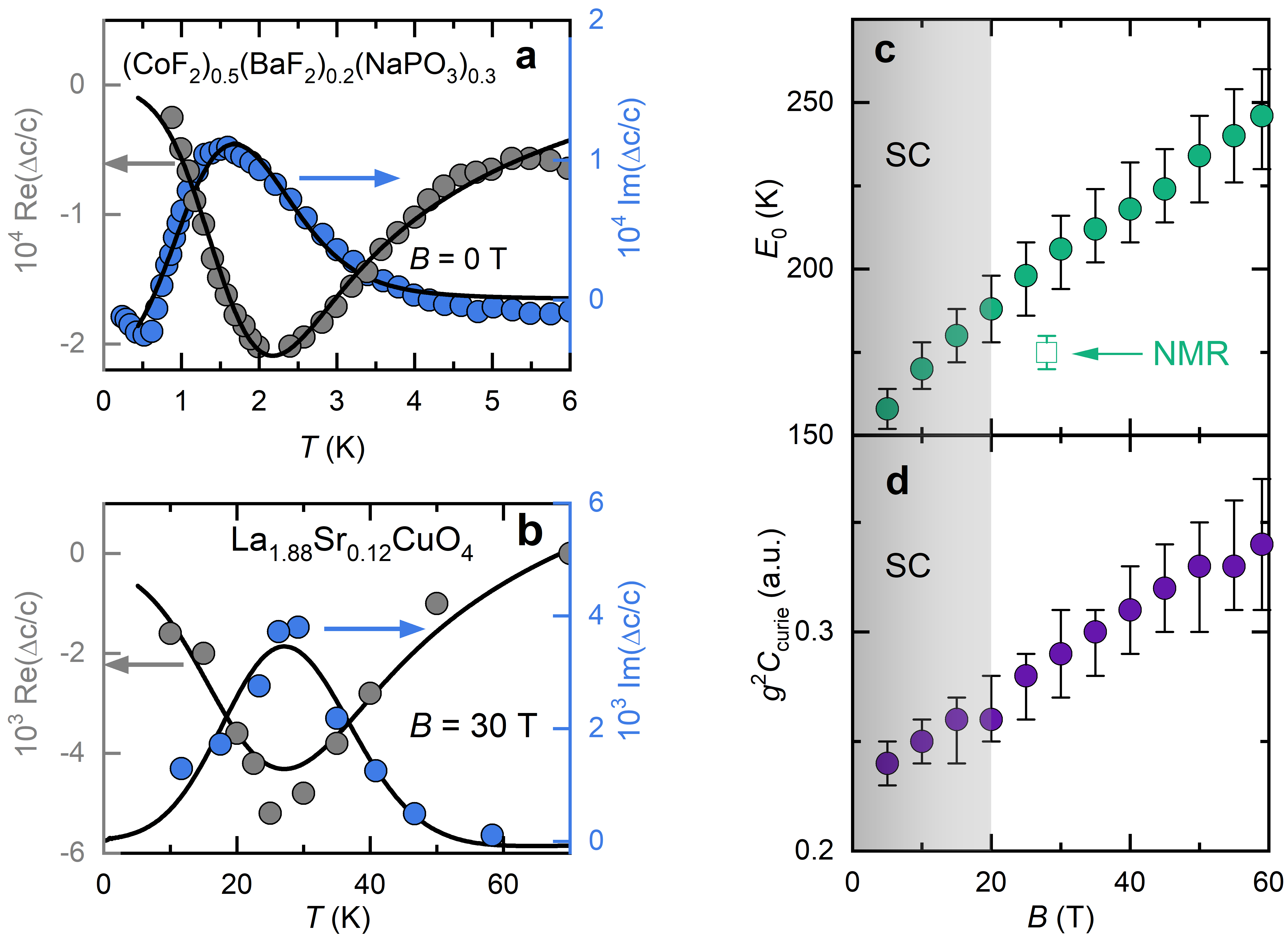}
\caption{Fitting of the experimental $\Re \textsf{e} \left( \Delta c/c \right)$ (gray circles, left scale) and $\Im \textsf{m} \left( \Delta c/c \right)$ (blue circles, right scale) for (a) the cobalt fluorophosphorate canonical spin glass in a longitudinal acoustic mode reproduced from Ref.\citep{Doussineau_EPL_1987} and (b) LSCO $p=0.12$ at $B=30$~T in the $(c_{11}-c_{12})/2$ transverse mode. In panels (a) and (b) a background sound velocity has been subtracted. The black lines are fit to the data using the dynamical susceptibility model. Fitting parameters for cobalt fluorophosphate are $E_0 \approx 18~$K and $g^2 C_{\rm curie} \approx 10^{-3}~$a.u.. Fitting parameters for LSCO $p=0.12$ are shown in panels (c) $E_0$ and (d) $g^2 C_{\rm{curie}}$. The other two free parameters, $g^2 \chi_0 \approx -4.5~$a.u. and $\Delta E_0 \approx 70~$K, don't show any significant field dependence. For comparison $E_0$ extracted from NMR $1/T_1$ at 28 T found in Fig. \ref{fig4} (empty square) is also shown. The error bars on a single parameter are estimated by monitoring the error of the fit with all other parameters fixed.}
\label{fig5}
\end{figure*}

A broad sound velocity minimum at $T_{\rm min}$ \citep{Doussineau_EPL_1987, Hsu_87, Hawkins_JAP_78} and an attenuation peak at $T_{\rm \alpha} \leq T_{\rm min}$ \citep{Doussineau_EPL_1987, Gaganidze_96} are common characteristics of - insulating or metallic - canonical spin glasses around the spin freezing temperature.  The sound velocity and attenuation of a cobalt fluorophosphorate spin glass shown in Fig. \ref{fig5}(a) exemplify those features. The comparison with the elastic properties of LSCO $p=0.12$ in the $(c_{11}-c_{12})/2$ mode shown in Fig. \ref{fig5}(b) is striking: both systems show remarkably similar phenomenology.  


In the following, we focus on the transverse $(c_{11}-c_{12})/2$ acoustic mode shown in Fig. \ref{fig3} and demonstrate that it can be semi-quantitatively reproduced by a phenomenological model developed for spin glasses. The strong increase observed in $c_{11}(T)$ at low temperature is not explained by this model and will be discussed later.

We use the phenomenological dynamical susceptibility model developed by Doussineau \textit{et al}. \citep{Doussineau_EPL_1987, Doussineau_ZPHYS_88}. Sound velocity and attenuation are expressed in terms of a complex elastic constant, $c$($\omega, T)$:

\begin{equation}
\label{Eq_bil}
c(\omega,T) = c_0 \left[ 1- g^2 \chi_4(\omega, T) \right] 
\end{equation}
With $c_0$ the bare elastic constant, $g$ the spin-phonon coupling constant and $\omega$ the ultrasound measurement frequency. Ultrasound quantities are deduced through:

\begin{equation}
\Delta v/v = \frac{1}{2} \Re \textsf{e} \left( \Delta c/c \right)
\end{equation}

\begin{equation}
\label{Eq_alpha}
\Delta \alpha (\textsf{dB/cm} ) =  \frac{\omega}{v} \frac{10}{\log(10)} \Im \textsf{m} \left( \Delta c/c \right)
\end{equation}

Here $\chi_4(\omega, T)$ is a dynamical susceptibility defined as:

\begin{equation}
\label{Eq_Suscept}
\chi_4(\omega, T) = \int \dfrac{d \tau_4}{\tau_4}  \dfrac{\chi_4(\omega = 0, T)}{1+i\omega \tau_4}
\end{equation}

$\chi_4(\omega=0, T)$ is a static susceptibility and $\tau_4 (T)$ is the correlation time of the spin fluctuations. In our case, since $S=1/2$, the magneto-acoustic coupling arises from the Waller mechanism (also called exchange-striction mechanism), \textit{i.e.} a modulation of the exchange interaction by the strain \citep{Luthi_book}. Consequently, the associated susceptibility is quadrupolar and the correlation time is involved in a four-spin correlation function. In contrast, the $1/T_1$ NMR relaxation rate is governed by a correlation time $\tau_2(T)$ which is involved in a two-spin correlation function. This can produce slight differences between $\Delta \alpha$ and $1/T_1$ in Fig. \ref{fig4} \citep{Volkmann_PRL_1986}. We use the following expressions for $\tau_4(T)$ and $\chi_4(\omega=0,T)$:

\begin{equation}
\label{Eq_tau}
\tau_4 (T) = \tau_{\infty} \exp(E_0/T)
\end{equation}

\begin{equation}
\chi_4(\omega=0,T) = \chi_0 + \frac{C_{\rm{curie}}}{T}
\label{Eq_chi}
\end{equation}

$C_{\rm{curie}}$ controls the amplitude of the lattice softening for $T\geq T_{\min}$, $E_0$ is an energy scale that governs $T_{\rm{min}}$ and $T_{\rm \alpha}$, $\chi_0$ is the constant term of the susceptibility and finally $\tau_{\infty}$ is the correlation time of spin fluctuations for $T \gg E_0$. Note that Eq. \ref{Eq_tau} and Eq. \ref{Eq_chi} are motivated by an analysis of $^{139}$La NMR $1/T_1$ \citep{Curro_PRL_00,Mitrovic_PRB_08} and ac-susceptibility measurements in the AFM glass of LSCO \citep{Wakimoto_PRB_00, Chou_PRL_95} respectively. As inferred from various experiments \citep{Markiewicz_PRB_01}, the value of $\tau_{\infty}$ is fixed to $\exp(-30) \simeq 10^{-13} $~s. Moreover, as usually done in spin glasses \citep{Doussineau_EPL_1987, Lundgren_JMMM_83}, and especially in the AFM glass phase of LSCO \citep{Curro_PRL_00,Markiewicz_PRB_01}, we consider that $\tau_4(T)$ is inhomogeneous using a gaussian-distribution of $E_0$ with full width at half maximum 2$\Delta E_0$. Within this framework it is possible to fit simultaneously $\Delta v/v$ and $\Delta \alpha$, and to extract both $E_0$ and $g^2 C_{\rm{curie}}$ as a function of magnetic field. A representative example is shown on Fig. \ref{fig5}(b): the model reproduces most of the salient features seen in the two ultrasound quantities.

 The evolution of the fitting parameters is shown in Fig. \ref{fig5}(c, d). Up to $B=60$ T - \textit{i.e.} well above our $T \rightarrow 0$ extrapolation of the vortex melting field $B_{\textrm v}$ on Fig. \ref{fig2}(c) - $E_0$ and $g^2 C_{\rm{curie}}$ increase continuously. The former increase is related to the non saturating values of the temperature scales $T_{\rm \alpha}$ and $T_{\rm min}$. The latter is explained by the continuous increase of the amplitudes of the lattice softening and attenuation peak up to 60 T (see Fig. \ref{fig3}).

The NMR $1/T_1$ data at $B=28$ T shown in Fig. \ref{fig4} can be fitted with the BPP formula using Eq. \ref{Eq_tau} for $\tau_2(T)$ and a gaussian distribution of activation energy $E_0$ \citep{Curro_PRL_00,Igor_thesis}. This parametrization of $1/T_1$ data yields an activation energy in fair agreement with $E_0$ inferred from ultrasound data (see Fig. \ref{fig5}(c)). It has been suggested previously that the activation energy is analogous to the spin-stiffness $2\pi\rho_s$ \citep{Teitel_PRB_00,Hunt_PRB_01, Igor_thesis}. The value of $E_0 \approx 200~$K found here for $B=20~$T is comparable to what is obtained in Nd-LSCO $x=0.12$ in zero magnetic field \citep{Tranquada_PRB_99,Hunt_PRB_01,Teitel_PRB_00}. It is an order of magnitude smaller than the spin stiffness of the antiferromagnetic parent compound La$_2$CuO$_4$ where $2\pi\rho_s\approx J$ \citep{Keimer_PRB_92}.

Finally, in the paramagnetic state of a classical Néel AFM $C_{\rm{curie}} \propto \mu^2$, where $\mu$ is the magnetic moment. Since the dynamical susceptibility model is purely phenomenological, we cannot extract microscopic information. As such, the increase of $g^2 C_{\rm{curie}}$ with magnetic field (see Fig. \ref{fig5}(d)) could originate from an enhanced $\mu$ \citep{Chang_PRB_2008, Lake_Nature_02} or from an increased magnetic volume \citep{Savici_PRB_02}.

\section{Discussion}

Let us summarize our results so far. \textbf{(i)} The $(c_{11}-c_{12})/2$ and $c_{11}$ modes show a softening for $T\geq T_{\rm{min}}$ and a hardening for $T\leq T_{\rm{ min}}$. Those features are enhanced by magnetic field and survive when superconductivity is strongly suppressed by the field. Consequently, neither feature is caused by superconductivity. We attribute this broad sound velocity minimum to the freezing of the AFM glass. \textbf{(ii)} The striking similarity of the ultrasound attenuation with the NMR relaxation rate $1/T_1$ shows that the AFM glass is also causing the anomalous attenuation peak in high magnetic field. \textbf{(iii)} The behaviour of the $(c_{11}-c_{12})/2$ elastic constant found in LSCO $p=0.12$ in high magnetic field is remarkably similar to what is found in canonical spin glasses. A dynamical susceptibility model, developed in the context of spin glasses, reproduces all features of the anomalous ultrasound properties in the $(c_{11}-c_{12})/2$ mode. 

The similar decrease of $T_{\rm min}$ and $T_{\rm c}$ with magnetic field $B\leq 14$ T in LSCO at $p\approx 0.14$ has previously motivated a scenario in which a competing lattice instability - that produces a
lattice softening for $T>T_{\rm c}$  - is quenched by the onset of
superconductivity that induces a hardening for $T<T_{\rm c}$
\citep{Nohara_PRL_93}. While we observe the same behaviour in LSCO $p=0.12$ for $B\leq 2$T (see Appendix \ref{appendix_superconductivity} for more details), this scenario does not hold at higher field where we observe an increase of $T_{\rm min}$.
All measurements reported here in LSCO $p=0.12$ support the interpretation that the ultrasound anomalies are caused by the AFM glass phase via spin-phonon coupling \citep{Frachet_19}. 


In the following we discuss some implications of the aforementioned results. In particular we comment on the magnetic field effect on the ultrasound properties, the relation of this study with previous elastic experiments and the symmetry of the AFM quasi-static fluctuations.


\subsection{The special coupling with $B_{\rm 1g}$ strain}

In canonical spin glasses such as cobalt fluorophosphorate, the magnetic moments are frozen in a random manner (note however that metallic spin glass CuMn show short ranged ($\sim 20~\angstrom$) SDW correlations \citep{mydosh}). Consequently, longitudinal and transverse acoustic modes couple similarly to the spins in such systems (see Ref. \citep{Doussineau_ZPHYS_88}). The magnetic moments of LSCO have similar dynamical properties as canonical spin glasses: they gradually freeze as the system is cooled down, such that the onset temperature depends on the probe frequency \citep{MHJ_review_LSCO, Romer_PRB_13}. However, the moments in LSCO arrange in a pattern displaying incommensurate AFM character, and Bragg peaks indicating correlation lengths as high as $\sim 200~ \angstrom$ in LSCO $x=0.12$ are observed in neutron diffraction experiments \citep{Yamada_PRB_98, Fujita_PRB_02, Wakimoto_PRB_00_Neutrons, Chang_PRB_2008}. Consequently in LSCO the coupling between the frozen spins and the lattice varies dramatically from one mode to another, as shown in Fig. \ref{fig1}. The anomalous softening for $T \geq T_{\rm min}$ is observed only in modes transforming according to the $B_{\rm 1g}$ irreducible representation (see Table \ref{table_elastic_constant} and Fig. \ref{fig1}). Note however, that we cannot exclude a similar coupling of the AFM glass to $B_{\rm 2g}$ mode. Nonetheless this suggests a special role of the $B_{\rm 1g}$ mode. 

Within the framework of the dynamical susceptibility model, the lattice softening in the $B_{1g}$ mode is caused by the growth of a Curie-like susceptibility $\chi_4(\omega=0,T)$. Eq. \ref{Eq_bil} is reminiscent of the elastic constant $c=\textrm{d}^{2}F/\textrm{d}\epsilon^{2}$ calculated using a Landau free energy $F$ containing a bilinear coupling $F_{\rm c}=g\epsilon Q$ \citep{Luthi_book}, with $\epsilon$ a strain and $Q$ an order parameter. Indeed, within such a model, the softening is directly related to the increasing mean-field susceptibility of $Q$, $\Delta v/v \propto - g^2\chi_Q$. For this bilinear coupling to exist, both $\epsilon$ and $Q$ must transform according to the same irreducible representation.  
In this context, our result would suggest that the order parameter (and the fluctuations) associated with the AFM glass has a $B_{1g}$, \textit{i.e.} nematic, character.

\begin{figure}
\includegraphics[width=8.6cm]{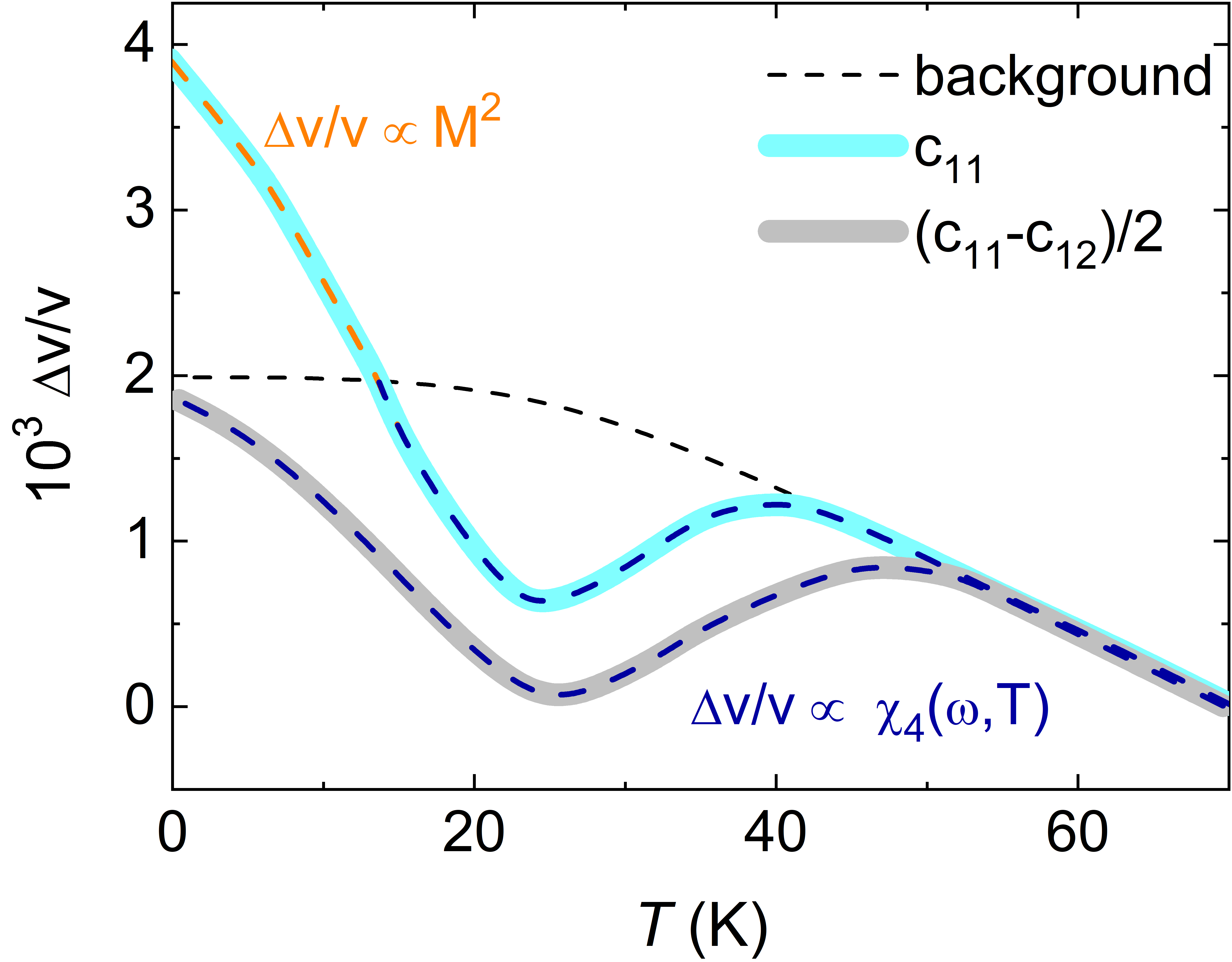}
\caption{Schematic representation of the sound velocity of the $(c_{11}-c_{12})/2$ (grey) and $c_{11}$ (cyan) modes in LSCO $p=0.12$. Black dashed line is the background elastic constant arising from anharmonicity of ionic potential \citep{Varshni_PRB_70}. $(c_{11}-c_{12})/2$ can be fully reproduced by the dynamical susceptibility model involving $\chi_4(\omega,T)$ (dashed blue line). For $T\rightarrow 0$, the difference between the sound velocity of the $(c_{11}-c_{12})/2$ mode and its background tends to zero. This contrasts with the behaviour of $c_{11}$. For $T\rightarrow 0$, the sound velocity of the $c_{11}$ mode is larger than the background sound velocity. This means that in addition to the contribution from $\chi_4(\omega,T)$ (dashed blue line) that causes the minimum in $c_{11}$, another component contributes at low $T$. As discussed in the text, a contribution proportional to $M ^2$ (dashed orange line) could explain the rapid increase of $c_{11}$ at low temperature.}
\label{fig8}
\end{figure}



Although conjectural in the absence of a measurement of the $B_{\rm 2g}$ mode, this interpretation of the ultrasound data is evocative of the $B_{1g}$ susceptibility observed by symmetry-resolved Raman scattering in LSCO at $x=0.10$ \citep{Tassini_PRL_05}. It is consistent with evidence of charge and spin stripe orders in this compound \citep{Wakimoto_PRB_00_Neutrons,Ando_PRL_02,Croft_PRB_14, Choi_ArxiV_2020}. Nematicity can indeed result from fluctuating stripes \citep{Kivelson_Nature_98}. We note that, at this doping, the observed $B_{1g}$ susceptibility develops well below the pseudogap temperature $T^\star$. Indeed, our ultrasound experiment does not detect any significant $B_{1g}$ susceptibility in the vicinity of $T^\star\approx 130~$K \citep{Choiniere_PRB_18}. The lack of $B_{1g}$ susceptibility at the pseudogap temperature is also reported in symmetry-resolved electronic Raman scattering experiments in Bi$_2$Sr$_2$CaCu$_2$O$_{8+\delta}$~ \citep{Auvray_Ncomm_19}. The onset temperature of our detection of $B_{1g}$ susceptibility is actually comparable to the CDW onset temperature $T_{\rm CDW} = 70 \pm 15$ K \citep{Croft_PRB_14,Christensen_14,Thampy_PRB_14}. This suggests that, in LSCO $p=0.12$, charge-stripe order  triggers slow magnetic fluctuations \citep{Baeck_PRB17,Arsenault_PRB18,Julien_99} with nematic character.

\begin{figure}
\centering
\includegraphics[width=8.6cm]{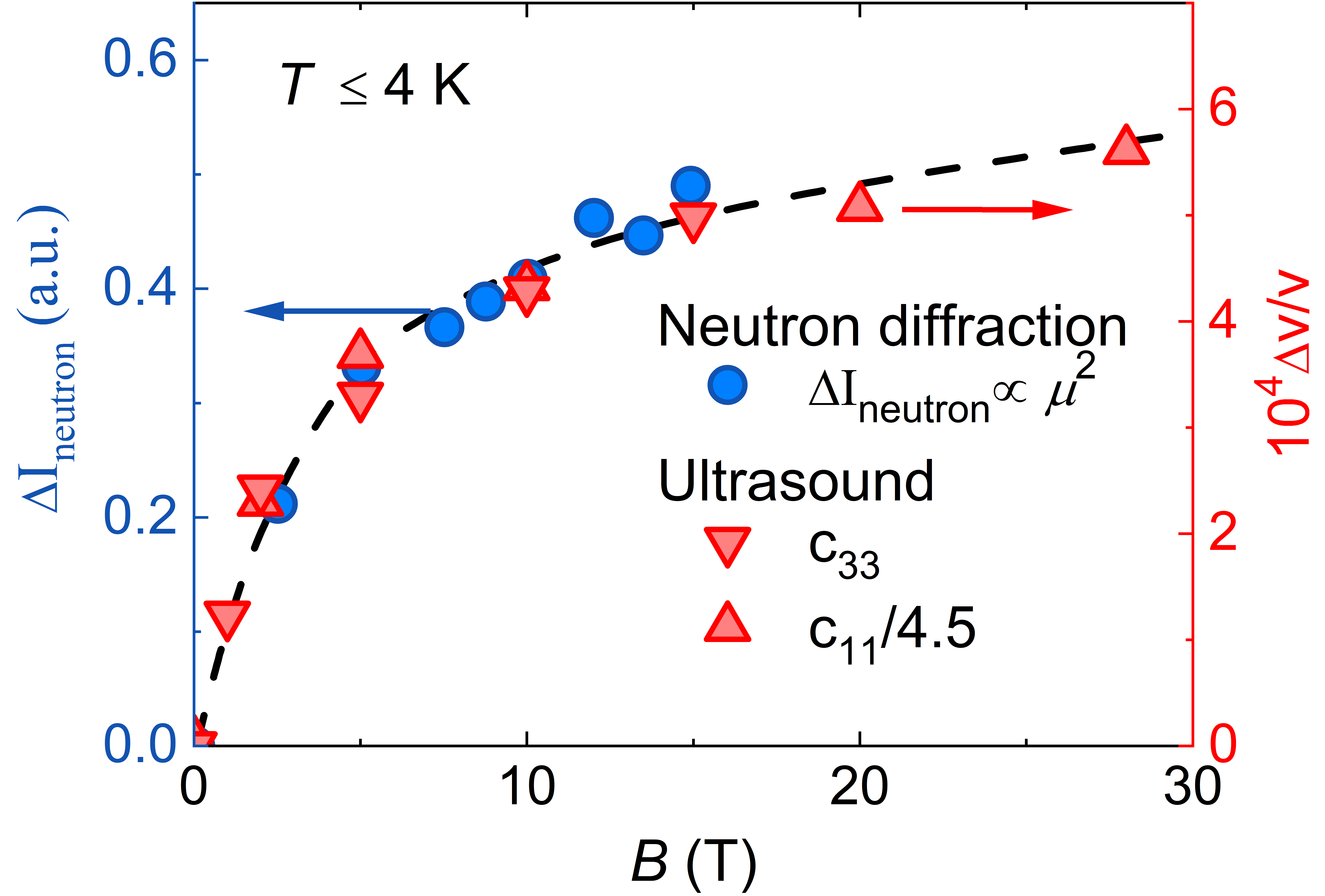}
\caption{Comparison between the magnetic field dependence of the superlattice Bragg peak intensity of incommensurate AFM seen by neutron diffraction, $\Delta I _ {\rm neutron } $ (blue circles, left scale), and the sound velocity, $\Delta v /v$, in the $c_{11}$ and $c_{33}$ acoustic modes (up and down red triangles respectively, right scale). The neutron diffraction intensity is reproduced from Ref. \citep{Chang_PRB_2008}. The sound velocity in the $c_{11}$ mode is divided by a factor 4.5. The sound velocity measurements presented here are taken at $T \leq 4~$K in field-cooled conditions. The dashed line is a guide to the eye.}
\label{fig7}
\end{figure}

\subsection{The effect of the magnetic field}
\label{discussion_magnetic_field_effect}

Previous neutron scattering and $\mu$SR experiments have shown that the magnetism of LSCO at $p \approx 0.12$ is enhanced by a magnetic field  \citep{Katano_PRB_00, Lake_Nature_02, Savici_PRL_05, Khaykovich_PRB05, Chang_PRL_07, Chang_PRB_2008}, and this effect has been ascribed to a competition between superconducting and AFM order parameters. In line with this interpretation, we observe that the ultrasound signatures of the AFM glass are strengthened by a magnetic field (see Fig. \ref{fig2}, \ref{fig3} and \ref{fig5}). 
The magnetic field dependence of the ultrasound properties does not saturate up to 60 T and the magnetic field induced softening appears at temperatures as high as $T \approx 50~$K (see Fig. \ref{fig3}). These observations are puzzling since at this doping $T_{\rm c} \approx 29~$K and the extrapolation of the vortex melting line leads to $B_{\rm v}(T \longrightarrow 0) \approx 20~$T. This raises important questions on the effect of magnetic fields on the magnetic freezing and the possible resilience of superconducting fluctuations in high field.

We note that this behaviour is reminiscent of the magnetoresistance producing an upturn in the resistivity of superconducting LSCO in high fields. This magnetoresistance is observed up to $T \simeq 100$ K at the doping level $p=0.12$ \citep{Boebinger_PRL_1996}. The spin freezing has been previously discussed as a cause of the resistivity upturn in La-based cuprates \citep{Harshman_PRB88,Hayden_91,Watanabe_JPSJ96,Sun_PRL03,Bourgeois_Arxiv_2019}. Consequently, it is possible that the large magnetoresistance observed in LSCO $p=0.12$ above $T_{\rm c}$ is related to the field-induced gradual slowing down of magnetic fluctuations observed here. 


\subsection{Differences between $c_{11}$ and $(c_{11}-c_{12})/2$}

Finally, we discuss the differences between the $c_{11}$ and $(c_{11}-c_{12})/2$ modes. As discussed above, the strength of the magneto-acoustic coupling is largest in the $(c_{11}-c_{12})/2$ mode, where the largest softening is observed (see Fig. \ref{fig1}). The second difference between the response in these two modes is the field-enhanced hardening that is seen in $c_{11}$ at low temperature. 
The situation is schematically depicted in Fig. \ref{fig8}. In the $(c_{11}-c_{12})/2$ mode, the difference between the measured sound velocity in the $T \rightarrow 0$ limit and the background velocity is negligibly small. On the other hand this difference is significant in the $c_{11}$ mode, with the measured sound velocity being larger than the background velocity. This behaviour echoes the results from previous studies performed on polycrystals. Earlier ultrasound studies suggested that this behaviour can be caused by LTT distortions \citep{Qu_APL_2006,Fukase_JJAP_1992}. On the other hand, anelastic experiments implied a coupling between strain and AFM glass domain wall motion \citep{Cordero_PRB_00}.Those studies are discussed in greater details in Appendix \ref{appendix_previous}.

Here we propose an alternative mechanism that could cause the low temperature stiffening in the $c_{11}$ mode. We start by noticing that a stiffening is also observed in the $c_{33}$ mode for $T\lesssim T_{\alpha}$ (see Fig. \ref{fig1} ). As shown in Fig. \ref{fig7}, the low temperature increase of the sound velocity in both the $c_{11}$ and $c_{33}$ modes has a field dependence that scales with the increase in $\mu^2$, the ordered moment squared inferred from neutron diffraction experiment, as discussed previously \citep{Prieur_04, Qu_APL_2006}. This scaling can be explained by invoking a biquadratic coupling $F_{\rm c}=\lambda\epsilon^2 M^2$ with $M$ the magnetization, and $\lambda$ a coupling constant. Note that $F_{\rm c}$ is symmetry-allowed for all elastic constants of Table \ref{table_elastic_constant}. This coupling produces $\Delta v/v \propto \lambda M^2$ and can naturally account for the experimental observations if we assume that the coupling constant $\lambda$ is larger for longitudinal modes ($c_{11}$ and $c_{33}$) than it is for transverse modes ($c_{44}$ and $(c_{11}-c_{12})/2$).


\section{Summary}
In summary, we studied sound velocity and attenuation in La$_{1.88}$Sr$_{0.12}$CuO$_4$ in high magnetic field. The behaviour of the $c_{11}$ and $(c_{11}-c_{12})/2$ elastic constants is highly anomalous. By comparing the anomalies with $^{139}$La NMR $1/T_{1}$ we confirm that they originate from the AFM glass phase via a magneto-acoustic coupling. A semi-quantitative analysis of this contribution is made based on a phenomenological model of spin glass systems. Our ultrasound data points toward a strong competition between spin freezing and superconductivity in high magnetic field. A symmetry analysis reveals that the slowing down of spin fluctuations could be associated with a growing nematic susceptibility.

\section*{ACKNOWLEDGMENTS}
We thank A. Böhmer, I. Paul and D. Campbell for fruitful discussions, as well as D. Destraz and O. Ivashko for their assistance with sample preparation.
Part of this work was performed at the LNCMI, a member of the European Magnetic Field Laboratory (EMFL). Work at the LNCMI was supported by the Laboratoire d’Excellence LANEF (ANR-10-LABX-51-01), French Agence Nationale de la Recherche (ANR) grant ANR-19-CE30-0019-01 (Neptun) and EUR grant NanoX n°ANR‐17‐EURE‐0009. Work in Zürich was supported by the Swiss National Science Foundation.
\bibliography{bibliography}

\begin{thebibliography}{10}

\bibitem{Tranquada_Nature_1995}
J.~M. Tranquada, B.~J. Sternlieb, J.~D. Axe, Y.~Nakamura, and S.~Uchida.
\newblock Evidence for stripe correlations of spins and holes in copper oxide
  superconductors.
\newblock {\em Nature}, 375:561--563, 1995.

\bibitem{Bozin_PRB_99}
E.~S. Bo\ifmmode~\check{z}\else \v{z}\fi{}in, S.~J.~L. Billinge, G.~H. Kwei,
  and H.~Takagi.
\newblock Charge-stripe ordering from local octahedral tilts: Underdoped and
  superconducting
  {${\mathrm{La}}_{2\ensuremath{-}x}{\mathrm{Sr}}_{x}{\mathrm{CuO}}_{4}$}$(0 <
  x < 0.30)$.
\newblock {\em Phys. Rev. B}, 59:4445--4454, Feb 1999.

\bibitem{Bianconi_PRL_1996}
A.~Bianconi, N.~L. Saini, A.~Lanzara, M.~Missori, T.~Rossetti, H.~Oyanagi,
  H.~Yamaguchi, K.~Oka, and T.~Ito.
\newblock Determination of the local lattice distortions in the
  {Cu${\mathrm{O}}_{2}$} plane of
  {${\mathrm{La}}_{1.85}{\mathrm{Sr}}_{0.15}{\mathrm{CuO}}_{4}$}.
\newblock {\em Phys. Rev. Lett.}, 76:3412--3415, 1996.

\bibitem{Saini_PRB_1997}
N.~L. Saini, A.~Lanzara, H.~Oyanagi, H.~Yamaguchi, K.~Oka, T.~Ito, and
  A.~Bianconi.
\newblock Local lattice instability and stripes in the {${\mathrm{CuO}}_{2}$}
  plane of the
  {${\mathrm{La}}_{1.85}$${\mathrm{Sr}}_{0.15}$${\mathrm{CuO}}_{4}$} system by
  polarized {XANES} and {EXAFS}.
\newblock {\em Phys. Rev. B}, 55:12759--12769, 1997.

\bibitem{Thampy_PRB_14}
V.~Thampy, M.~P.~M. Dean, N.~B. Christensen, L.~Steinke, Z.~Islam, M.~Oda,
  M.~Ido, N.~Momono, S.~B. Wilkins, and J.~P. Hill.
\newblock Rotated stripe order and its competition with superconductivity in
  {${\mathrm{La}}_{1.88}{\mathrm{Sr}}_{0.12}{\mathrm{CuO}}_{4}$}.
\newblock {\em Phys. Rev. B}, 90:100510, 2014.

\bibitem{Croft_PRB_14}
T.~P. Croft, C.~Lester, M.~S. Senn, A.~Bombardi, and S.~M. Hayden.
\newblock Charge density wave fluctuations in
  {${\text{La}}_{2\ensuremath{-}x}$${\text{Sr}}_{x}$${\text{CuO}}_{4}$} and
  their competition with superconductivity.
\newblock {\em Phys. Rev. B}, 89:224513, 2014.

\bibitem{Christensen_14}
N.~B.~Christensen et~al.
\newblock Bulk charge stripe order competing with superconductivity in
  {${\mathrm{La}}_{2\ensuremath{-}x}{\mathrm{Sr}}_{x}{\mathrm{CuO}}_{4}$}
  ($x=0.12$).
\newblock {\em arXiv}, 1404.3192, 2014.

\bibitem{Wen_Arxiv_CDW_LSCO}
J.~J. {Wen}, H.~{Huang}, S.~J. {Lee}, H.~{Jang}, J.~{Knight}, Y.~S. {Lee},
  M.~{Fujita}, K.~M. {Suzuki}, S.~{Asano}, S.~A. {Kivelson}, C.~C. {Kao}, and
  J.~S. {Lee}.
\newblock {Observation of two types of charge-density-wave orders in
  superconducting
  {${\text{La}}_{2\ensuremath{-}x}$${\text{Sr}}_{x}$${\text{CuO}}_{4}$}}.
\newblock {\em Nature Communications}, 10:3269, 2019.

\bibitem{MHJ_review_LSCO}
M.-H. Julien.
\newblock Magnetic order and superconductivity in
  {${\mathrm{La}}_{2-x}{\mathrm{Sr}}_{\mathit{x}}{\mathrm{CuO}}_{4}$}: a
  review.
\newblock {\em Physica B: Condensed Matter}, 329-333:693 -- 696, 2003.
\newblock Proceedings of the 23rd International Conference on Low Temperature
  Physics.

\bibitem{Romer_PRB_13}
A.~T. R\o{}mer, J.~Chang, N.~B. Christensen, B.~M. Andersen, K.~Lefmann,
  L.~M\"ahler, J.~Gavilano, R.~Gilardi, Ch. Niedermayer, H.~M. R\o{}nnow,
  A.~Schneidewind, P.~Link, M.~Oda, M.~Ido, N.~Momono, and J.~Mesot.
\newblock Glassy low-energy spin fluctuations and anisotropy gap in
  {${\mathrm{La}}_{2-x}{\mathrm{Sr}}_{\mathit{x}}{\mathrm{CuO}}_{4}$}.
\newblock {\em Phys. Rev. B}, 87:144513, 2013.

\bibitem{Suzuki_PRB_98}
T.~Suzuki, T.~Goto, K.~Chiba, T.~Shinoda, T.~Fukase, H.~Kimura, K.~Yamada,
  M.~Ohashi, and Y.~Yamaguchi.
\newblock Observation of modulated magnetic long-range order in
  {${\mathrm{La}}_{1.88}{\mathrm{Sr}}_{0.12}{\mathrm{CuO}}_{4}$}.
\newblock {\em Phys. Rev. B}, 57:R3229--R3232, 1998.

\bibitem{Nohara_PRL_93}
M.~Nohara, T.~Suzuki, Y.~Maeno, T.~Fujita, I.~Tanaka, and H.~Kojima.
\newblock Interplay between lattice softening and high-${T}_{\mathrm{c}}$
  superconductivity in
  {${\mathrm{La}}_{1.86}$${\mathrm{Sr}}_{0.14}$${\mathrm{CuO}}_{4}$}.
\newblock {\em Phys. Rev. Lett.}, 70:3447--3450, 1993.

\bibitem{Qu_APL_2006}
J.~F. Qu, Y.~Q. Zhang, X.~L. Lu, X.~Q. Xiang, Y.~L. Liao, G.~Li, and X.~G. Li.
\newblock Ultrasonic study on magnetic-field-induced stripe order in
  {${\mathrm{La}}_{1.88-x}{\mathrm{Sr}}_{0.12}{\mathrm{Ba}}_{x}{\mathrm{CuO}}_{4}$}.
\newblock {\em Applied Physics Letters}, 89(16):162508, 2006.

\bibitem{Qu_MSE_2006}
J.F. Qu, Y.Q. Zhang, X.Q. Xiang, X.L. Lu, and X.G. Li.
\newblock Effect of magnetic field on the charge-stripe phase in
  {${\mathrm{La}}_{2-x-y}{\mathrm{Nd}}_{\mathit{y}}{\mathrm{Sr}}_{\mathit{x}}{\mathrm{CuO}}_{4}$}:
  An ultrasonic attenuation study.
\newblock {\em Materials Science and Engineering: A}, 442(1):216 -- 219, 2006.
\newblock Proceedings of the 14th International Conference on Internal Friction
  and Mechanical Spectroscopy.

\bibitem{Hanaguri_PhysicaB_94}
T.~Hanaguri, T.~Fukase, T.~Suzuki, I.~Tanaka, and H.~Kojima.
\newblock Elastic anomalies in a
  {${\mathrm{La}}_{1.85}{\mathrm{Sr}}_{0.15}{\mathrm{CuO}}_{4}$} single crystal
  under high magnetic fields.
\newblock {\em Physica B: Condensed Matter}, 194-196:1579 -- 1580, 1994.

\bibitem{Sakita_PhysicaB_96}
S.~Sakita, T.~Suzuki, F.~Nakamura, M.~Nohara, Y.~Maeno, and T.~Fujita.
\newblock Elastic softening in single-crystalline
  {${\mathrm{La}}_{2\ensuremath{-}x}{\mathrm{Sr}}_{x}{\mathrm{CuO}}_{4}$}
  around x=1/8.
\newblock {\em Physica B: Condensed Matter}, 219-220:216 -- 218, 1996.
\newblock PHONONS 95.

\bibitem{Prieur_04}
Jean-Yves Prieur and Jacques Joffrin.
\newblock Ultrasonic velocity variations in
  {${\mathrm{La}}_{2\ensuremath{-}x}{\mathrm{Sr}}_{x}{\mathrm{CuO}}_{4}$}
  single crystals.
\newblock {\em physica status solidi (c)}, 1(11):3061--3064, 2004.

\bibitem{Frachet_19}
M.~Frachet, I.~Vinograd, R.~Zhou, S.~Benhabib, S.~Wu, H.~Mayaffre, S.~Krämer,
  S.~K. Ramakrishna, A.~Reyes, J.~Debray, T.~Kurosawa, N.~Momono, M.~Oda,
  S.~Komiya, S.~Ono, M.~Horio, J.~Chang, C.~Proust, D.~LeBoeuf, and M.-H.
  Julien.
\newblock Hidden magnetism at the pseudogap critical point of a high
  temperature superconductor.
\newblock {\em Nature Physics}, 16:1064, 2020.

\bibitem{Luthi_book}
B.~Luthi.
\newblock Physical acoustics in the solid state.
\newblock 2005.

\bibitem{Varshni_PRB_70}
Y.~P. Varshni.
\newblock Temperature dependence of the elastic constants.
\newblock {\em Phys. Rev. B}, 2:3952--3958, 1970.

\bibitem{Nohara_PRB_95}
M.~Nohara, T.~Suzuki, Y.~Maeno, T.~Fujita, I.~Tanaka, and H.~Kojima.
\newblock Unconventional lattice stiffening in superconducting
  {${\mathrm{La}}_{2\mathrm{\ensuremath{-}}\mathit{x}}$${\mathrm{Sr}}_{\mathit{x}}$${\mathrm{CuO}}_{4}$}
  single crystals.
\newblock {\em Phys. Rev. B}, 52:570--580, 1995.

\bibitem{Gugenberger_PRB_94}
F.~Gugenberger, C.~Meingast, G.~Roth, K.~Grube, V.~Breit, T.~Weber, H.~W\"uhl,
  S.~Uchida, and Y.~Nakamura.
\newblock Uniaxial pressure dependence of {${\mathit{T}}_{\mathit{c}}$} from
  high-resolution dilatometry of untwinned
  {${\mathrm{La}}_{2\mathrm{\ensuremath{-}}\mathit{x}}$${\mathrm{Sr}}_{\mathit{x}}$${\mathrm{CuO}}_{4}$}
  single crystals.
\newblock {\em Phys. Rev. B}, 49:13137--13142, May 1994.

\bibitem{Mitrovic_PRB_08}
V.~F. Mitrovi\ifmmode~\acute{c}\else \'{c}\fi{}, M.-H. Julien, C.~de~Vaulx,
  M.~Horvati\ifmmode~\acute{c}\else \'{c}\fi{}, C.~Berthier, T.~Suzuki, and
  K.~Yamada.
\newblock Similar glassy features in the $^{139}\text{L}\text{a}$
  {$\mathrm{NMR}$} response of pure and disordered
  {${\text{La}}_{1.88}{\text{Sr}}_{0.12}{\text{CuO}}_{4}$}.
\newblock {\em Phys. Rev. B}, 78:014504, 2008.

\bibitem{Hunt_PRB_01}
A.~W. Hunt, P.~M. Singer, A.~F. Cederstr\"om, and T.~Imai.
\newblock Glassy slowing of stripe modulation in
  {${(\mathrm{L}\mathrm{a},\mathrm{E}\mathrm{u},\mathrm{N}\mathrm{d})}_{2\ensuremath{-}x}({\mathrm{S}\mathrm{r},\mathrm{B}\mathrm{a})}_{x}{\mathrm{CuO}}_{4}:
  \mathrm{A}{ }^{63}\mathrm{Cu}$ and ${}^{139}\mathrm{La}$ {NQR} study down to
  350 {mK}}.
\newblock {\em Phys. Rev. B}, 64:134525, Sep 2001.

\bibitem{Simovic_PRB_03}
B.~Simovi\ifmmode~\check{c}\else \v{c}\fi{}, P.~C. Hammel, M.~H\"ucker,
  B.~B\"uchner, and A.~Revcolevschi.
\newblock Experimental evidence for a glass forming stripe liquid in the
  magnetic ground state of
  {${\mathrm{La}}_{1.65}{\mathrm{Eu}}_{0.2}{\mathrm{Sr}}_{0.15}{\mathrm{CuO}}_{4}$}.
\newblock {\em Phys. Rev. B}, 68:012415, Jul 2003.

\bibitem{Curro_PRL_00}
N.~J. Curro, P.~C. Hammel, B.~J. Suh, M.~H\"ucker, B.~B\"uchner, U.~Ammerahl,
  and A.~Revcolevschi.
\newblock Inhomogeneous low frequency spin dynamics in
  {${\mathrm{La}}_{1.65}{\mathrm{Eu}}_{0.2}{\mathrm{Sr}}_{0.15}{\mathrm{CuO}}_{4}$}.
\newblock {\em Phys. Rev. Lett.}, 85:642--645, Jul 2000.

\bibitem{BPP_PR_1948}
N.~Bloembergen, E.~M. Purcell, and R.~V. Pound.
\newblock Relaxation effects in nuclear magnetic resonance absorption.
\newblock {\em Phys. Rev.}, 73:679--712, 1948.

\bibitem{Doussineau_EPL_1987}
P.~Doussineau, A.~Levelut, M.~Matecki, J.~P. Renard, and W.~Schön.
\newblock Acoustic and magnetic studies of an insulating spin glass.
\newblock {\em Europhysics Letters ({EPL})}, 3(2):251--258, 1987.

\bibitem{Hsu_87}
T.~C. Hsu and J.~B. Marston.
\newblock Measurement of ultrasound velocity in the spin-glass
  {$\mathrm{CuMn}$}.
\newblock {\em Journal of Applied Physics}, 61(5):2074--2077, 1987.

\bibitem{Hawkins_JAP_78}
G.~F. Hawkins and R.~L. Thomas.
\newblock Ultrasonic studies of spin glasses: {CuMn}.
\newblock {\em Journal of Applied Physics}, 49(3):1627--1629, 1978.

\bibitem{Gaganidze_96}
E.~Gaganidze and P.~Esquinazi.
\newblock Temperature dependence of the sound attenuation at the spin glass
  transition of metallic spin glasses.
\newblock {\em Czechoslovak Journal of Physics}, 46:2227--2228, 1996.

\bibitem{Doussineau_ZPHYS_88}
P.~Doussineau, A.~Levelut, and W.~Schön.
\newblock Acoustic susceptibility of an insulating spin-glass.
\newblock {\em Zeitschrift für Physik B Condensed Matter}, 73:89--102, 1988.

\bibitem{Volkmann_PRL_1986}
U.~G. Volkmann, R.~B\"ohmer, A.~Loidl, K.~Knorr, U.~T. H\"ochli, and
  S.~Hauss\"uhl.
\newblock Dipolar and quadrupolar freezing in
  ${(\mathrm{KBr})}_{1\ensuremath{-}x}{(\mathrm{KCN})}_{x}$.
\newblock {\em Phys. Rev. Lett.}, 56:1716--1719, Apr 1986.

\bibitem{Wakimoto_PRB_00}
S.~Wakimoto, S.~Ueki, Y.~Endoh, and K.~Yamada.
\newblock Systematic study of short-range antiferromagnetic order and the
  spin-glass state in lightly doped
  {${\mathrm{La}}_{2\ensuremath{-}x}{\mathrm{Sr}}_{x}{\mathrm{CuO}}_{4}$}.
\newblock {\em Phys. Rev. B}, 62:3547--3553, 2000.

\bibitem{Chou_PRL_95}
F.~C. Chou, N.~R. Belk, M.~A. Kastner, R.~J. Birgeneau, and Amnon Aharony.
\newblock Spin-glass behavior in
  {${\mathrm{La}}_{1.96}{\mathrm{Sr}}_{0.04}{\mathrm{CuO}}_{4}$}.
\newblock {\em Phys. Rev. Lett.}, 75:2204--2207, 1995.

\bibitem{Markiewicz_PRB_01}
R.~S. Markiewicz, F.~Cordero, A.~Paolone, and R.~Cantelli.
\newblock Cluster spin-glass distribution functions in
  {${\mathrm{La}}_{2\ensuremath{-}x}{\mathrm{Sr}}_{x}{\mathrm{CuO}}_{4}$}.
\newblock {\em Phys. Rev. B}, 64:054409, Jul 2001.

\bibitem{Lundgren_JMMM_83}
L.~Lundgren, P.~Svedlindh, and O.~Beckman.
\newblock Anomalous time dependence of the susceptibility in a
  {$\mathrm{Cu(Mn)}$} spin glass.
\newblock {\em Journal of Magnetism and Magnetic Materials}, 31-34:1349 --
  1350, 1983.

\bibitem{Igor_thesis}
I.~Vinograd.
\newblock Études par résonance magnétique nucléaire des ordres en
  compétitions dans les cuprates supraconducteurs.
\newblock {\em PhD Thesis}, Universit\'e Grenoble Alpes, 2019.

\bibitem{Teitel_PRB_00}
G.~B. Teitel'baum, I.~M. Abu-Shiekah, O.~Bakharev, H.~B. Brom, and J.~Zaanen.
\newblock Spin dynamics and ordering of a cuprate stripe antiferromagnet.
\newblock {\em Phys. Rev. B}, 63:020507, Dec 2000.

\bibitem{Tranquada_PRB_99}
J.~M. Tranquada, N.~Ichikawa, and S.~Uchida.
\newblock Glassy nature of stripe ordering in
  {${\mathrm{La}}_{1.6\ensuremath{-}x}{\mathrm{Nd}}_{0.4}{\mathrm{Sr}}_{x}{\mathrm{CuO}}_{4}$}.
\newblock {\em Phys. Rev. B}, 59:14712--14722, Jun 1999.

\bibitem{Keimer_PRB_92}
B.~Keimer, N.~Belk, R.~J. Birgeneau, A.~Cassanho, C.~Y. Chen, M.~Greven, M.~A.
  Kastner, A.~Aharony, Y.~Endoh, R.~W. Erwin, and G.~Shirane.
\newblock Magnetic excitations in pure, lightly doped, and weakly metallic
  {${\mathrm{La}}_{2}{\mathrm{CuO}}_{4}$}.
\newblock {\em Phys. Rev. B}, 46:14034--14053, Dec 1992.

\bibitem{Chang_PRB_2008}
J.~Chang, Ch. Niedermayer, R.~Gilardi, N.~B. Christensen, H.~M. R\o{}nnow,
  D.~F. McMorrow, M.~Ay, J.~Stahn, O.~Sobolev, A.~Hiess, S.~Pailhes, C.~Baines,
  N.~Momono, M.~Oda, M.~Ido, and J.~Mesot.
\newblock Tuning competing orders in
  {${\text{La}}_{2\ensuremath{-}x}{\text{Sr}}_{x}{\text{CuO}}_{4}$} cuprate
  superconductors by the application of an external magnetic field.
\newblock {\em Phys. Rev. B}, 78:104525, 2008.

\bibitem{Lake_Nature_02}
B.~Lake, H.~M. Rønnow, N.~B. Christensen, G.~Aeppli, K.~Lefmann, D.~F.
  McMorrow, P.~Vorderwisch, P.~Smeididl, N.~Mangkorntong, T.~Sasagawa,
  M.~Nohara, H.~Takagi, and T.~E. Mason.
\newblock Antiferromagnetic order induced by an applied magnetic field in a
  high-temperature superconductor.
\newblock {\em Nature}, 415:299--302, 2002.

\bibitem{Savici_PRB_02}
A.~T. Savici, Y.~Fudamoto, I.~M. Gat, T.~Ito, M.~I. Larkin, Y.~J. Uemura, G.~M.
  Luke, K.~M. Kojima, Y.~S. Lee, M.~A. Kastner, R.~J. Birgeneau, and K.~Yamada.
\newblock Muon spin relaxation studies of incommensurate magnetism and
  superconductivity in stage-4 {${\mathrm{La}}_{2}{\mathrm{CuO}}_{4.11}$} and
  {${\mathrm{La}}_{1.88}{\mathrm{Sr}}_{0.12}{\mathrm{CuO}}_{4}$}.
\newblock {\em Phys. Rev. B}, 66:014524, Jul 2002.

\bibitem{mydosh}
J.~A. Mydosh.
\newblock {\em Spin Glasses: an experimental introduction}.
\newblock Taylor and Francis, 1993.

\bibitem{Yamada_PRB_98}
K.~Yamada, C.~H. Lee, K.~Kurahashi, J.~Wada, S.~Wakimoto, S.~Ueki, H.~Kimura,
  Y.~Endoh, S.~Hosoya, G.~Shirane, R.~J. Birgeneau, M.~Greven, M.~A. Kastner,
  and Y.~J. Kim.
\newblock Doping dependence of the spatially modulated dynamical spin
  correlations and the superconducting-transition temperature in
  {${\mathrm{La}}_{2\mathrm{\ensuremath{-}}\mathit{x}}{\mathrm{Sr}}_{x}{\mathrm{CuO}}_{4}$}.
\newblock {\em Phys. Rev. B}, 57:6165--6172, 1998.

\bibitem{Fujita_PRB_02}
M.~Fujita, K.~Yamada, H.~Hiraka, P.~M. Gehring, S.~H. Lee, S.~Wakimoto, and
  G.~Shirane.
\newblock Static magnetic correlations near the insulating-superconducting
  phase boundary in
  {${\mathrm{La}}_{2\ensuremath{-}x}{\mathrm{Sr}}_{x}{\mathrm{CuO}}_{4}$}.
\newblock {\em Phys. Rev. B}, 65:064505, 2002.

\bibitem{Wakimoto_PRB_00_Neutrons}
S.~Wakimoto, R.~J. Birgeneau, M.~A. Kastner, Y.~S. Lee, R.~Erwin, P.~M.
  Gehring, S.~H. Lee, M.~Fujita, K.~Yamada, Y.~Endoh, K.~Hirota, and
  G.~Shirane.
\newblock Direct observation of a one-dimensional static spin modulation in
  insulating {${\mathrm{La}}_{1.95}{\mathrm{Sr}}_{0.05}{\mathrm{CuO}}_{4}$}.
\newblock {\em Phys. Rev. B}, 61:3699--3706, 2000.

\bibitem{Tassini_PRL_05}
L.~Tassini, F.~Venturini, Q.-M. Zhang, R.~Hackl, N.~Kikugawa, and T.~Fujita.
\newblock Dynamical properties of charged stripes in
  {${\mathrm{La}}_{2\ensuremath{-}x}{\mathrm{Sr}}_{x}{\mathrm{CuO}}_{4}$}.
\newblock {\em Phys. Rev. Lett.}, 95:117002, Sep 2005.

\bibitem{Ando_PRL_02}
Yoichi Ando, Kouji Segawa, Seiki Komiya, and A.~N. Lavrov.
\newblock Electrical resistivity anisotropy from self-organized one
  dimensionality in high-temperature superconductors.
\newblock {\em Phys. Rev. Lett.}, 88:137005, Mar 2002.

\bibitem{Choi_ArxiV_2020}
J.~Choi, Q.~Wang, S.~Jöhr, N.~B. Christensen, J.~Küspert, D.~Bucher,
  D.~Biscette, M.~Hücker, T.~Kurosawa, N.~Momono, M.~Oda, O.~Ivashko,
  M.~v.~Zimmermann, M.~Janoschek, and J.~Chang.
\newblock Disentangling intertwined quantum states in a prototypical cuprate
  superconductor.
\newblock arXiv:2009.06967 2020.

\bibitem{Kivelson_Nature_98}
S.~A. Kivelson, E.~Fradkin, and V.~J. Emery.
\newblock Electronic liquid-crystal phases of a doped {Mott} insulator.
\newblock {\em Nature}, 393:550, 1998.

\bibitem{Choiniere_PRB_18}
O.~Cyr-Choini\`ere, R.~Daou, F.~Lalibert\'e, C.~Collignon, S.~Badoux,
  D.~LeBoeuf, J.~Chang, B.~J. Ramshaw, D.~A. Bonn, W.~N. Hardy, R.~Liang, J.-Q.
  Yan, J.-G. Cheng, J.-S. Zhou, J.~B. Goodenough, S.~Pyon, T.~Takayama,
  H.~Takagi, N.~Doiron-Leyraud, and Louis Taillefer.
\newblock Pseudogap temperature ${T}^{*}$ of cuprate superconductors from the
  {Nernst} effect.
\newblock {\em Phys. Rev. B}, 97:064502, Feb 2018.

\bibitem{Auvray_Ncomm_19}
N.~Auvray, B.~Loret, S.~Benhabib, M.~Cazayous, R.~D. Zhong, J.~Scheeloch, G.~D.
  Gu, A.~Forget, D.~Colson, Paul. I, A.~Sacuto, and Y.~Gallais.
\newblock Nematic fluctuations in the cuprate superconductor
  {${\text{Bi}}_{2}{\text{Sr}}_{2}{\text{CaCu}}_{2}{\text{O}}_{8+\delta}$}.
\newblock {\em Nature Communications}, 10(5209), 2019.

\bibitem{Baeck_PRB17}
S.-H. Baek, A.~Erb, and B.~B\"uchner.
\newblock Low-energy spin dynamics and critical hole concentrations in
  ${\mathrm{la}}_{2\ensuremath{-}x}{\mathrm{sr}}_{x}{\mathrm{cuo}}_{4}$
  $(0.07\ensuremath{\le}x\ensuremath{\le}0.2)$ revealed by $^{139}\mathrm{La}$
  and $^{63}\mathrm{Cu}$ nuclear magnetic resonance.
\newblock {\em Phys. Rev. B}, 96:094519, Sep 2017.

\bibitem{Arsenault_PRB18}
A.~Arsenault, S.~K. Takahashi, T.~Imai, W.~He, Y.~S. Lee, and M.~Fujita.
\newblock $^{139}\mathrm{La}$ nmr investigation of the charge and spin order in
  a ${\mathrm{la}}_{1.885}{\mathrm{sr}}_{0.115}{\mathrm{cuo}}_{4}$ single
  crystal.
\newblock {\em Phys. Rev. B}, 97:064511, Feb 2018.

\bibitem{Julien_99}
M.-H. Julien, F.~Borsa, P.~Carretta, M.~Horvati\ifmmode~\acute{c}\else
  \'{c}\fi{}, C.~Berthier, and C.~T. Lin.
\newblock Charge segregation, cluster spin glass, and superconductivity in
  ${\mathrm{la}}_{1.94}{\mathrm{sr}}_{0.06}{\mathrm{cuo}}_{4}$.
\newblock {\em Phys. Rev. Lett.}, 83:604--607, Jul 1999.

\bibitem{Katano_PRB_00}
S.~Katano, M.~Sato, K.~Yamada, T.~Suzuki, and T.~Fukase.
\newblock Enhancement of static antiferromagnetic correlations by magnetic
  field in a superconductor
  {${\mathrm{La}}_{2\ensuremath{-}x}{\mathrm{Sr}}_{x}{\mathrm{CuO}}_{4}$} with
  $x=0.12$.
\newblock {\em Phys. Rev. B}, 62:R14677--R14680, 2000.

\bibitem{Savici_PRL_05}
A.~T. Savici, A.~Fukaya, I.~M. Gat-Malureanu, T.~Ito, P.~L. Russo, Y.~J.
  Uemura, C.~R. Wiebe, P.~P. Kyriakou, G.~J. MacDougall, M.~T. Rovers, G.~M.
  Luke, K.~M. Kojima, M.~Goto, S.~Uchida, R.~Kadono, K.~Yamada, S.~Tajima,
  T.~Masui, H.~Eisaki, N.~Kaneko, M.~Greven, and G.~D. Gu.
\newblock Muon spin relaxation studies of magnetic-field-induced effects in
  high-${T}_{c}$ superconductors.
\newblock {\em Phys. Rev. Lett.}, 95:157001, Oct 2005.

\bibitem{Khaykovich_PRB05}
B.~Khaykovich, S.~Wakimoto, R.~J. Birgeneau, M.~A. Kastner, Y.~S. Lee,
  P.~Smeibidl, P.~Vorderwisch, and K.~Yamada.
\newblock Field-induced transition between magnetically disordered and ordered
  phases in underdoped
  ${\mathrm{la}}_{2\ensuremath{-}x}{\mathrm{sr}}_{x}\mathrm{Cu}{\mathrm{o}}_{4}$.
\newblock {\em Phys. Rev. B}, 71:220508, Jun 2005.

\bibitem{Chang_PRL_07}
J.~Chang, A.~P. Schnyder, R.~Gilardi, H.~M. R\o{}nnow, S.~Pailhes, N.~B.
  Christensen, Ch. Niedermayer, D.~F. McMorrow, A.~Hiess, A.~Stunault,
  M.~Enderle, B.~Lake, O.~Sobolev, N.~Momono, M.~Oda, M.~Ido, C.~Mudry, and
  J.~Mesot.
\newblock Magnetic-field-induced spin excitations and renormalized spin gap of
  the underdoped
  {${\mathrm{La}}_{1.895}{\mathrm{Sr}}_{0.105}{\mathrm{CuO}}_{4}$}
  superconductor.
\newblock {\em Phys. Rev. Lett.}, 98:077004, Feb 2007.

\bibitem{Boebinger_PRL_1996}
G.~S. Boebinger, Yoichi Ando, A.~Passner, T.~Kimura, M.~Okuya, J.~Shimoyama,
  K.~Kishio, K.~Tamasaku, N.~Ichikawa, and S.~Uchida.
\newblock Insulator-to-metal crossover in the normal state of
  {${\mathrm{La}}_{2\ensuremath{-}\mathit{x}}{\mathrm{Sr}}_{\mathit{x}}{\mathrm{CuO}}_{4}$}
  near optimum doping.
\newblock {\em Phys. Rev. Lett.}, 77:5417--5420, Dec 1996.

\bibitem{Harshman_PRB88}
D.~R. Harshman, G.~Aeppli, G.~P. Espinosa, A.~S. Cooper, J.~P. Remeika, E.~J.
  Ansaldo, T.~M. Riseman, D.~Ll. Williams, D.~R. Noakes, B.~Ellman, and T.~F.
  Rosenbaum.
\newblock Freezing of spin and charge in
  ${\mathrm{la}}_{2\ensuremath{-}x}{\mathrm{sr}}_{x}\mathrm{Cu}{\mathrm{o}}_{4}$.
\newblock {\em Phys. Rev. B}, 38:852--855, Jul 1988.

\bibitem{Hayden_91}
S.~M. Hayden, G.~Aeppli, H.~Mook, D.~Rytz, M.~F. Hundley, and Z.~Fisk.
\newblock Magnetic fluctuations in
  ${\mathrm{la}}_{1.95}$${\mathrm{ba}}_{0.05}$${\mathrm{cuo}}_{4}$.
\newblock {\em Phys. Rev. Lett.}, 66:821--824, Feb 1991.

\bibitem{Watanabe_JPSJ96}
Isao Watanabe, Tadashi Adachi, Satoshi Yairi, Yoji Koike, and Kanetada
  Nagamine.
\newblock Change of the dynamics of internal fields in the normal state of
  la2-xsrxcuo4 observed by muon-spin-relaxation.
\newblock {\em Journal of the Physical Society of Japan}, 77(12):124716, 2008.

\bibitem{Sun_PRL03}
X.~F. Sun, Seiki Komiya, J.~Takeya, and Yoichi Ando.
\newblock Magnetic-field-induced localization of quasiparticles in underdoped
  ${\mathrm{l}\mathrm{a}}_{2\ensuremath{-}x}{\mathrm{s}\mathrm{r}}_{x}{\mathrm{c}\mathrm{u}\mathrm{o}}_{4}$
  single crystals.
\newblock {\em Phys. Rev. Lett.}, 90:117004, Mar 2003.

\bibitem{Bourgeois_Arxiv_2019}
P.~Bourgeois-Hope, S.~Y. Li, F.~Laliberté, S.~Badoux, S.~M. Hayden, N.~Momono,
  T.~Kurosawa, K.~Yamada, H.~Takagi, Nicolas Doiron-Leyraud, and Louis
  Taillefer.
\newblock Link between magnetism and resistivity upturn in cuprates: a thermal
  conductivity study of
  {${\mathrm{La}}_{2\ensuremath{-}\mathit{x}}{\mathrm{Sr}}_{\mathit{x}}{\mathrm{CuO}}_{4}$}.
\newblock arXiv:1910.08126 2019.

\bibitem{Fukase_JJAP_1992}
T.~Fukase, T.~Hanaguri, T.~Nomoto, T.~Goto, Y.~Koike, Shinohara T., T.~Sato,
  I.~Tanaka, and H.~Kojima.
\newblock Ultrasonic and {NQR} studies of structural phase transitions and
  superconductivity in
  {${\mathrm{La}}_{2-x}$${\mathrm{Sr}}_{x}$${\mathrm{CuO}}_{4}$}.
\newblock {\em JJAP series}, 7:213--218, 1992.

\bibitem{Cordero_PRB_00}
F.~Cordero, A.~Paolone, R.~Cantelli, and M.~Ferretti.
\newblock Anelastic spectroscopy of the cluster spin-glass phase in
  {${\mathrm{La}}_{2\ensuremath{-}x}{\mathrm{Sr}}_{x}{\mathrm{CuO}}_{4}$}.
\newblock {\em Phys. Rev. B}, 62:5309--5312, Sep 2000.

\bibitem{Chang_PRB_12}
J.~Chang, J.~S. White, M.~Laver, C.~J. Bowell, S.~P. Brown, A.~T. Holmes,
  L.~Maechler, S.~Str\"assle, R.~Gilardi, S.~Gerber, T.~Kurosawa, N.~Momono,
  M.~Oda, M.~Ido, O.~J. Lipscombe, S.~M. Hayden, C.~D. Dewhurst, R.~Vavrin,
  J.~Gavilano, J.~Kohlbrecher, E.~M. Forgan, and J.~Mesot.
\newblock Spin density wave induced disordering of the vortex lattice in
  superconducting
  {${\mathrm{La}}_{2\ensuremath{-}x}{\mathrm{Sr}}_{x}{\mathrm{CuO}}_{4}$}.
\newblock {\em Phys. Rev. B}, 85:134520, Apr 2012.

\bibitem{Goto_PhysicaB_1998}
T.~Goto, T.~Suzuki, K.~Chiba, T.~Shinoda, M.~Mori, and T.~Fukase.
\newblock Field-induced spin reorientation and low-temperature structural phase
  transition in
  {${\mathrm{La}}_{2-x}{\mathrm{Sr}}_{\mathit{x}}{\mathrm{CuO}}_{4}$}.
\newblock {\em Physica B: Condensed Matter}, 246-247:572 -- 575, 1998.

\bibitem{Koyama_PRB_95}
Y.~Koyama, Y.~Wakabayashi, K.~Ito, and Y.~Inoue.
\newblock Low-temperature structural transitions and
  {${\mathit{T}}_{\mathit{c}}$} suppression in
  {${\mathrm{La}}_{2\mathrm{\ensuremath{-}}\mathit{x}}$${\mathit{M}}_{\mathit{x}}$${\mathrm{CuO}}_{4}$}
  {(M=Ba, Sr)}.
\newblock {\em Phys. Rev. B}, 51:9045--9051, Apr 1995.

\bibitem{Horibe_PRB_00}
Y.~Horibe, Y.~Inoue, and Y.~Koyama.
\newblock Direct observation of dynamic local structure in
  {${\mathrm{La}}_{2\ensuremath{-}x}{\mathrm{Sr}}_{x}{\mathrm{CuO}}_{4}$}
  around $x=0.12$.
\newblock {\em Phys. Rev. B}, 61:11922--11927, May 2000.

\bibitem{Cordero_PRB_01}
F.~Cordero, A.~Paolone, R.~Cantelli, and M.~Ferretti.
\newblock Pinning of the domain walls of the cluster spin-glass phase in the
  low-temperature-tetragonal phase of
  {${\mathrm{La}}_{2\ensuremath{-}x}{\mathrm{Ba}}_{x}{\mathrm{CuO}}_{4}$}.
\newblock {\em Phys. Rev. B}, 64:132501, Aug 2001.

\end{thebibliography}

\appendix
\renewcommand\thefigure{\thesection.\arabic{figure}} 
\setcounter{figure}{0}

\section{Vortex lattice contribution to the ultrasound properties}

\label{appendix_vortex}

In our experiments with $B \parallel$ [001], the $c_{11}$ acoustic mode probes the compression modulus of the vortex lattice (VL), whereas $(c_{11}-c_{12})/2$ probes its shear modulus. The latter is notorious for its small value, beyond our resolution. On the other hand the compression modulus of the VL should make a detectable contribution. This contribution should give a step-like increase of the sound velocity and an attenuation peak when decreasing temperature through the depinning transition of the vortex lattice. Both these features should move to lower temperatures with increasing field and disappear for $B \geq B_{\rm c2}$. 

In Fig. \ref{fig2}(a, b) there is no evidence of the vortex lattice depinning transition. It has a negligible contribution to the longitudinal sound velocity in LSCO $p=0.12$. This is due to the low value of the irreversibility field, and presumably also because of the two-dimensional and disordered character of the VL at this doping level  \citep{Chang_PRB_12}. Consequently no signature of the vortex lattice depinning transition is observed in Fig. \ref{fig2}(a, b) and Fig. \ref{fig3}.

\section{Impact of superconductivity on the ultrasound properties}

\label{appendix_superconductivity}

 We discuss the effect of superconductivity on the ultrasound properties which is best observed at low fields. Below 5 T or so, $T_{\rm{min}}$ coincides with $T_{\rm c}$ in both $c_{11}$ and $(c_{11}-c_{12})/2$ in our LSCO $p=0.12$ sample (see Fig. \ref{fig1} and Fig. \ref{fig2}). This behaviour has also been observed in LSCO $p \approx 0.14$ up to 14 T by Nohara and coworkers \citep{Nohara_PRL_93}.  In this field range $T_{\rm min}$ appears to be primarily set by superconductivity. 

In LSCO, superconductivity can induce an increase of the sound velocity for $T<T_{\rm c}$ via two mechanisms. First, the superconducting order parameter has a direct coupling with the lattice for $T<T_{\rm c}$. In cuprates, this coupling produces a hardening in the superconducting state, for both longitudinal and transverse modes \citep{Nohara_PRB_95}. Consequently, an upturn can occur at $T_{\rm c}$ in both $(c_{11}-c_{12})/2$ and $c_{11}$. 
The second possible mechanism is indirect, and involves the competition between superconductivity and magnetism. In zero and low field, the growth of the magnetic susceptibility $\chi_4(\omega=0,T)$, signaled by the softening of the sound velocity, can be tempered by the onset of superconductivity. If $\chi_4(\omega=0,T)$ is sufficiently modified through $T_{\rm c}$, it can result in an upturn at $T_{\rm c}$. As a result of these two possible mechanisms, at zero and low magnetic field, we observe $T_{\rm min} = T_{\rm c}$, and $T_{\rm min}$ decreases as field increases.

However, for $B \geq 5~$T, $T_{\rm min}$ increases with magnetic field, meaning that the mechanism causing the softening for $T>T_{\rm{min}}$ and the hardening for $T<T_{\rm{min}}$ observed for $B\geq 5$ T in LSCO $p=0.12$ does not involve the coupling of the superconducting order parameter to the lattice. Increasing the magnetic field above $B=14~$T at $p \approx 0.14$ leads to the same observation \citep{Frachet_19}. As field increases the superconducting contribution to the sound velocity becomes weaker and the spin freezing contribution larger. For $B\geq 5$~T the superconducting contribution is dwarfed by the contribution from the magnetic slowing down. This explains why the difference between $T_{\rm min}$  and $T_{\rm \alpha}=T_{\rm f}$ is large and strongly field dependent for $B<5$~T, while smaller and constant for higher fields (see Fig. \ref{fig2}(c)).

\begin{figure}
    \centering
    \includegraphics[width=8.6cm]{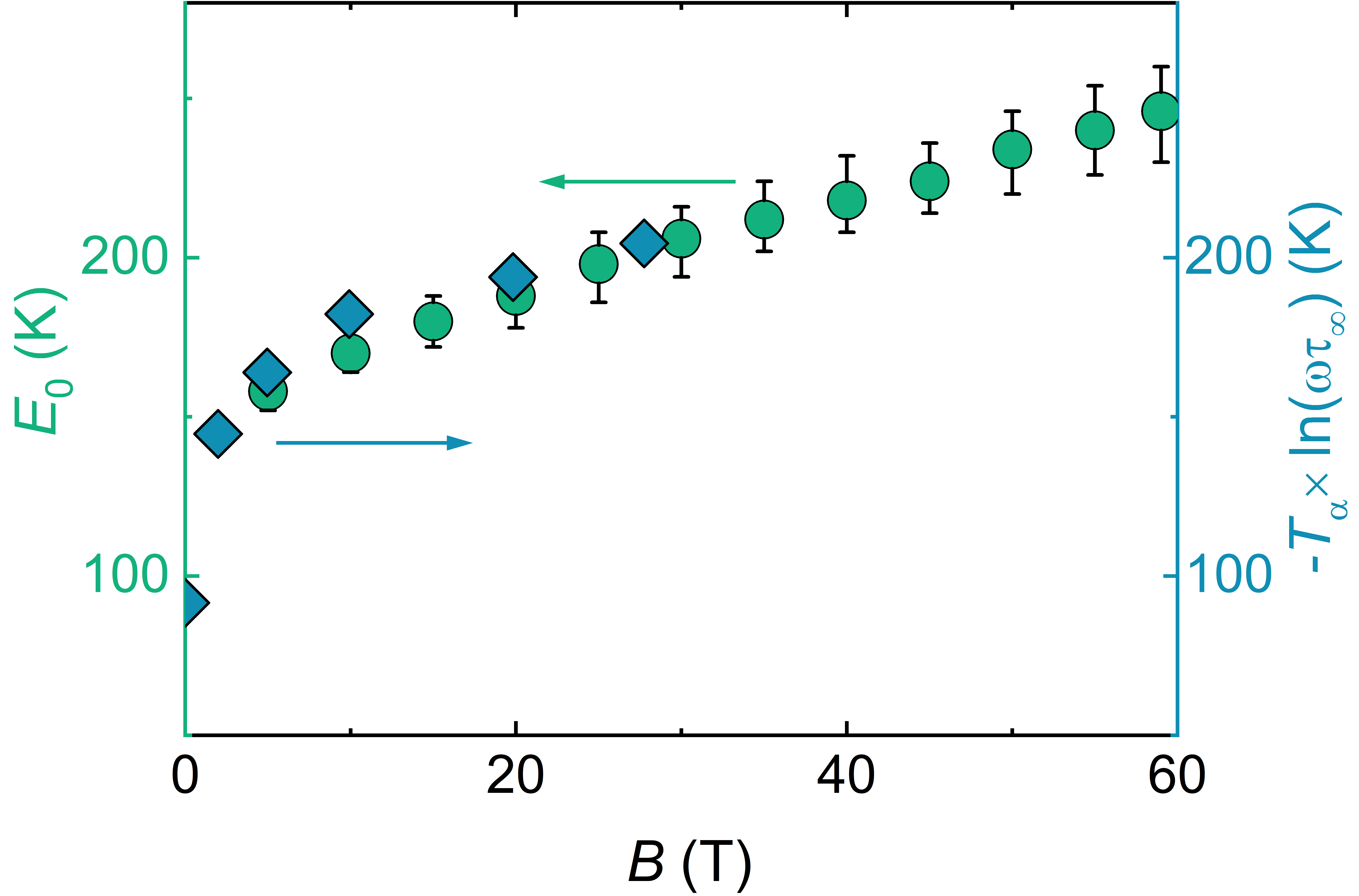}
    \caption{Comparison between the $E_0$ energy scale (circles, left scale) extracted from the fitting procedure of Fig. \ref{fig5} and $-T_{\rm{\alpha}}\times \ln(\omega \tau_{\infty})$ (diamonds, right scale) determined directly from the data shown in Fig. \ref{fig2}(b), $\omega = 2 \pi \times 110~$MHz and $\tau_{\infty} \approx 10^{-13}~$s. The energy scale $-T_{\rm{\alpha}}\times \ln(\omega \tau_{\infty})$ corresponds to an experimental determination of $E_0$ solely based on $T_{\rm \alpha}$, using  the condition $\omega \tau_4(T_{\rm \alpha}) =1$ in Eq. \ref{Eq_tau} (see text). The rapid drop of $-T_{\rm{\alpha}}\times \ln(\omega \tau_{\infty})$ at low $B$ is most likely due to the impact of superconductivity on the spin dynamics. Error bars on this quantity are smaller than the size of the symbols.}
    \label{fig6}
\end{figure}

The temperature scale $T_{\rm \alpha}$ is insensitive of a direct contribution from superconductivity. While in conventional superconductors, the opening of the superconducting gap causes an attenuation drop, there is no corresponding behaviour in LSCO $p \approx 0.12$. However $\Delta \alpha(T)$ can be indirectly impacted by the onset of superconductivity at low field because the latter modifies the spin dynamics that controls $\Delta \alpha(T)$. This is best illustrated by the zero field $\Delta \alpha(T)$ that shows a remarkable kink at $T_{c}$ and then a maximum at $T_{\rm \alpha}\simeq 9.5$~K (see Fig. \ref{fig2}(b)). Within the dynamical susceptibility model, the ultrasound attenuation is mostly governed by the energy scale $E_0$ entering $\tau_4$ as indicated by Eq. \ref{Eq_alpha}. The kink anomaly at $T_{\rm c}$ in $\Delta \alpha(T)$ in zero field can be interpreted as a decrease of $E_0$ for $T<T_{\rm c}$ caused by the onset of superconductivity. 

Because the dynamical susceptibility model does not take into account the impact of superconductivity on spin dynamics, we use an alternative scheme in order to extract $E_0$ for $B<5$ T. At $T=T_{\alpha}$, the condition $\omega \tau_4(T)=1$ is met. Solving Eq. \ref{Eq_tau} for $E_0$ at $T=T_{\alpha}$ hence yields $E_0=-T_{\alpha} / \ln(\omega\tau_{\infty})$ \citep{Igor_thesis}. In Fig. \ref{fig6} we compare this $T_{\alpha}$ derived $E_0$ with $E_0$ of Fig. \ref{fig5}(c) obtained with the parametrization of the ultrasound data. Good agreement is found between the two estimations of $E_0$. As seen in Fig. \ref{fig6}, $-T_{\alpha} / \ln(\omega\tau_{\infty})$ decreases rapidly at low fields, dropping from $E_0\sim 150 K$ for $B=5$ T to $E_0=92$ K for $B=0$. This rapid drop reflects the competition between spin freezing and superconductivity.\\

\section{Previous experiments on polycrystals}
\label{appendix_previous}
In section \ref{discussion_magnetic_field_effect} we discuss a scenario that would explain the lattice hardening seen mainly in the $c_{11}$ elastic constant at low temperature. Here, we detail previous putative explanations that have been proposed to explain a similar phenomenology from ultrasound and anelastic studies of polycrystalline La-based compounds.
 
We first consider previous ultrasound experiments. Around $p \approx 0.12$, in several La-based cuprates and in particular LSCO  \citep{Fukase_JJAP_1992, Qu_MSE_2006, Goto_PhysicaB_1998,Qu_APL_2006}, the sound velocity of longitudinal waves increases markedly at low temperature while an attenuation peak occurs. This behaviour is most likely related to the anomalous hardening of the $c_{11}$ elastic constant, as suggested by the similar temperature scales and field-enhancement. Interestingly, those previous ultrasound studies have shown that the temperature scale of the lattice hardening evolves smoothly from LBCO $p=0.12$ to LSCO $p=0.12$, and correspond to the coincident OMT-LTT and charge-stripe transition in the former \citep{Fukase_JJAP_1992,Qu_APL_2006}. Based on these experiments it has been proposed that in LSCO $p \approx 0.12$ the low temperature lattice hardening arises from the parallel development of local and/or fluctuating LTT distorsions and charge-stripes. However, in this scenario it is unclear why the lattice hardening is observed only at low temperature - comparable to $T_{\rm{\alpha}}$ - whereas, in LSCO $p\approx 0.12$, LTT-type tilts are found at temperatures as high as $T=100$~K in electron diffraction experiments \citep{Koyama_PRB_95,Horibe_PRB_00} and LTT-type reflections are observed up to the OMT-HTT transition temperature $T_{\rm st}$ in X-Ray diffraction \citep{Christensen_14}.

Now let us consider anelastic experiments in LSCO and LBCO polycrystals. These have shown that the Young modulus increases markedly at low temperature \citep{Cordero_PRB_01}. The elastic energy loss coefficient shows a step-like increase for $T\leq T_{\rm f}$ and a plateau down to the lowest $T$ \citep{Cordero_PRB_00}. Based on a comparison with nuclear quadrupolar resonance (NQR) experiment, as well as the effect of oxygen vacancy, it has been inferred that the anelastic anomalies arise from a strain induced motion of the antiferromagnetic domain walls of the AFM glass phase \citep{Cordero_PRB_00}. In particular this naturally explains why these anomalies are observed at temperatures of the order of $T_{\rm{\alpha}}$. However, these experiments are performed at much lower frequency than the ultrasound ones (within the kHz range) and thus don't necessarily probe the same relaxation process as ultrasound experiments \citep{Markiewicz_PRB_01}.

\end{document}